\documentclass[a4paper,11pt]{article}
\pdfoutput=1 

\usepackage{jheppub} 
\usepackage{ytableau} 
\usepackage{slashed} 
\usepackage{braket}
\usepackage{multirow}
\usepackage{xcolor}
\usepackage{caption}
\usepackage{float}
\usepackage{times}
\usepackage[enableskew]{youngtab} 
\usepackage{hyperref} 
\hypersetup{
    colorlinks=true,
    linkcolor=black,
    filecolor=magenta,      
    urlcolor=blue,
}
\usepackage[T1]{fontenc} 

\bibstyle{JHEP}

\newcommand{\Sec}{Section }

\title{\boldmath Light composite fermions from holography}

\author[a]{Raimond Abt,}
\author[a]{Johanna Erdmenger,}
\author[b]{Nick Evans,}
\author[b]{Konstantinos S. Rigatos}

\affiliation[a]{ Institut f{\"u}r Theoretische Physik und Astrophysik,
Julius-Maximilians-Universit{\"a}t W{\"u}rzburg,\\ Am Hubland, 97074 W{\"u}rzburg, Germany.}
\affiliation[b]{School  of  Physics  \&  Astronomy  and  STAG  Research  Centre,  University  of  Southampton,\\Highfield, Southampton  SO$17$ $1$BJ,  UK.}

\abstract{Motivated by Beyond the Standard Model theories of composite
  fermions or top partners, we propose a holographic mechanism that
  generates light baryonic states in a strongly coupled gauge theory.
  The starting point are the fermionic fluctuations of {\it massive}
  probe branes embedded into AdS$_5  \times S^5$. We first consider
  the D3/probe D7-brane system. We derive in detail the fermionic
  fluctuation equations and show the supersymmetric degeneracy of the
  mesinos with the mesons.  Here we view the fermionic mesinos as
  potential realizations of composite fermions or top partners. We then add higher dimension operators and study their impact on the mesino spectrum. In particular we show that the ground state mesino mass can  be pushed  to an arbitrarily light value by a suitable choice of the coupling of the higher dimension operator, $g$. No matter the value of $g$, the masses of higher excited states never fall below the mass of the ground state at $g=0$. We also present similar results for the supersymmetric D3/D3 and D3/D5 systems.}

\begin{document} 
\maketitle
\flushbottom
\newpage

\section{Introduction}

It has long been a matter of interest whether strongly coupled gauge theories can generate light or even massless fermionic bound states (baryons), since these might form the basis for a composite model of standard model fermions \cite{Dimopoulos:1980hn}. A related problem is
to generate the experimentally observed top quark mass in composite Higgs models, in which the Higgs particle is a pseudo-Nambu Goldstone boson. This requires baryonic top partners \cite{Kaplan:1991dc, Ferretti:2013kya} that are light relative to the typical hadronic scale in the strongly coupled sector. The AdS/CFT correspondence \cite{Maldacena:1997re, Witten:1998qj, Gubser:1998bc} has provided a new window on strongly interacting gauge dynamics that may potentially be useful as a new approach to Beyond the Standard Model (BSM) physics. It motivates us here to look afresh at a mechanism for generating light or massless baryons in top-down holographic models, in which the use of a top-down string theory D-brane construction provides control over the field content of the dual gauge theory. As a starting point for new BSM analyses, in this paper we begin by carefully fixing the details of top-down gauge/gravity duality models required for investigating fermionic modes. Somewhat removed from the phenomenological BSM models mentioned,
we  study a rigorously understood top-down construction of an ${\cal N}=2$ gauge theory with massive quarks. In this theory, the meson states and their supersymmetric partners, the mesinos, can be analytically computed and lie at a scale determined by the quark mass. Here we will determine how higher dimension operators may be used to generate abnormally light mesino states. There are two sets of mesino states corresponding to different representations of the supersymmetry algebra. One of them is very similar to a QCD baryon multiplet since the lowest mass entry in this multiplet consists of a product of three elementary fermion fields (of course a true baryon at large $N_c$ is made of $N_c$ quarks and must be represented by a baryon vertex in the dual \cite{Witten:1998xy}).  In the future, we hope to extend our mechanism to holographic descriptions of more phenomenlogically relevant gauge theories, in particular of theories displaying chiral symmetry breaking, extending the results of \cite{Babington:2003vm}.

The D3/probe D7-brane system \cite{Karch:2002sh, Kruczenski:2003be,
  Erdmenger:2007cm} provides a clean holographic description of a
strongly coupled gauge theory with quark matter for which an easily
calculable dual description exists. The gauge theory is an ${\cal
  N}=2$ supersymmetric theory with hypermultiplets added to the base
${\cal N}=4$ super Yang Mills theory. The gravity dual in the quenched
approximation consists of probe D7-branes embedded in AdS$_5\times
S^5$ space that wrap a subspace which asymptotically near the boundary is  AdS$_5 \times S^3$ \cite{Karch:2002sh}. The quark mass and condensate are explicitly present in the model as holographic modes and determine the near-boundary behaviour of the embedding functions. The meson spectrum, corresponding to fluctuations of the brane about their vacuum configuration, was computed in \cite{Kruczenski:2003be}. The fermionic spectrum in the {\it massless} theory was fully derived in \cite{Kirsch:2006he}. In the same paper, a phenomenological bottom-up rule was used  to guess the equations of motion for the fluctuations in the massive case, reproducing the expected spectrum, with further results in  \cite{Erdmenger:2007cm}. A full derivation of the equations of motion for the massive case has been completed in the unpublished notes \cite{informal}. The first task we set ourselves here is to provide an explicit derivation of these equations  and to check the supersymmetric degeneracy of the spectrum. The results in \cite{Faraggi:2011bb} are also a useful related reference.

The general strategy for studying fermionic open string fluctuations of probe D-branes is to write a ten-dimensional action for a 16-component Majorana-Weyl spinor and then perform the pull-back onto the world-volume of the D-brane \cite{Martucci:2005rb}. For supersymmetric embeddings such as the ones we consider, where the probe D-brane lies flat in the background space-time, this pull-back simply corresponds to dropping derivative terms for the bulk directions. In cases with a curved embedding, as required for modelling chiral symmetry breaking in non-supersymmetric backgrounds for instance, this would be a more involved process. However we do not consider such cases here. 
For the supersymmetric case we re-write the metric in terms of vielbeins, determine the non-vanishing spin-connection components of the background and evaluate the Dirac operator on the world-volume of the probe branes. In this way, we  obtain the individual terms that appear in the first-order equations of motion. These terms can be divided into one Dirac operator associated with the asymptotically AdS space and one on the transverse $S^3$ in the case of the D7-brane probe. The eigenvalues of the $S^3$ operator split into two sectors with an opposite sign. We need to consider both of these sectors and we choose to denote by ${\cal G}$ the set of modes derived from the positive sign of the spherical eigenvalues, while we use ${\cal F}$ to describe the modes associated with the negative eigenvalue on the sphere. Each of these generate distinct mesino states in the field theory. The usual holographic operator matching shows that the former set of modes is naively associated with bound states of two quarks and a gaugino, while the latter are quark-squark bound states.  

The first order Dirac equations can be squared to a Klein-Gordon second order form. A factor of $\gamma^\rho$ associated with the radial direction $\rho$  in AdS is still present in the second-order equation.  However, it is possible to write the solution in terms of eigenvectors of $\gamma^\rho$ with eigenvalues $\pm 1$. When the mesino is massive, the leading term from either solution in the UV region near the boundary  is associated with the field theory source $J$ and the sub-leading term with the operator ${\cal O}$. This identification requires some care, as we discuss based on the previous results of \cite{Laia:2011wf} . We present analytic solutions for the supersymmetric case and reconfirm earlier analysis \cite{Kirsch:2006he} that show the solutions match the expected supersymmetric spectrum. We also present a detailed numerical approach for determining the
mesino masses which we need for our later analysis including higher dimension operators.

So far, the model considered does not give rise to light baryons since the masses of the baryon-like mesinos are tied  to the meson spectrum by supersymmetry. To proceed, one possible addition to the theories are higher dimension operators. Witten's double trace prescription \cite{Witten:2001ua} allows such operators to be introduced easily as modifications of the UV boundary conditions on the holographic solutions. Previously this has been done in the D3/D7 system for Nambu-Jona-Lasinio type four-fermion operators in \cite{Evans:2016yas}. Here, instead, we consider adding operators of "mesino squared'' form which naively will generate a shift in the mesino mass in the effective description of the low-energy hadrons. We show that as the coupling of these operators is raised, the mesino masses can be driven to light values relative to the rest of the spectrum. For small values of the coupling of this operator, the shift in the mesino mass is small and linear, but above a  critical value of the coupling the shift in the mesino mass is suddenly sharp and much larger. Inspite of this, the mesino mass can only be pushed to zero for asymptotically large values of the coupling, presumably reflecting that fermionic states cannot become tachyonic and condense. Our approach provides at least one (tuned) mechanism for generating light composite fermions in strongly coupled gauge theories. We also study the radially excited states of the mesinos and show that their masses are bounded from below and do not become light along with the lowest state whose mass approaches zero. 

In later sections we also extend this derivation to the other supersymmetric systems D3/D3 and D3/D5 in order to test that the behaviour we see is generic which it seems to be. 

This paper is organized as follows: in \Sec \ref{d3d7sec} we present the background
of the D$3$/D$7$-brane system together with the mass spectrum of the mesons dual to bosonic fluctuations of the probe D$7$-brane derived in \cite{Kruczenski:2003be}. We discuss the fermionic fluctuations of the D$7$-brane in \Sec \ref{sec: fermionic fluctuations in D3/D7} where we show how to obtain the mass spectrum of the dual mesinos both analytically and numerically. In \Sec \ref{sec:intermediate} we then apply the methods discussed in \Sec \ref{sec: fermionic fluctuations in D3/D7} for the D$3$/D$7$ system to the D$3$/D$5$, and D$3$/D$3$.
We end this paper with concluding comments and remarks in \Sec \ref{sec:conclusions}. Our notational choices are discussed in the  appendix. 

\section{D3/D7-brane system: Background and bosonic fluctuations}
\label{d3d7sec}
We begin by reviewing the bosonic sector of the canonical D$3$/probe
D$7$-brane system which describes the ${\cal N}=4$ super Yang-Mills
theory with quenched ${\cal N}=2$ matter multiplets
\cite{Karch:2002sh, Kruczenski:2003be}. This is the example for which
we will study the fermionic fluctuations in full detail below.

\subsection{D3-Brane background geometry}
\label{dbrnssugrabckgrnd}

According to the AdS/CFT correspondence, a stack of  D$3$-branes generates  type IIB supergravity theory  on AdS$_5\times S^5$,
which is dual to 
${\cal N}=4$ gauge theory \cite{Maldacena:1997re, Witten:1998qj, Gubser:1998bc}.
We choose the following basis representation of the AdS$_5\times S^5$ metric,
\begin{align}
\label{eq: D3 background}
	ds^2
	=
	G_{MN}dx^Mdx^N
	=
	\frac{r^{2}}{R^2} \eta_{\mu\nu}dx^\mu dx^\nu + \frac{R^2}{r^2} d \rho^2  + \frac{R^2 \rho^2}{r^2} d \Omega_{3}^2 + \frac{R^2}{r^2} \delta_{\tilde{m} \tilde{n}} dw^{\tilde{m}} dw^{\tilde{n}}\,,
\end{align}
where $M,N=0,\dots 9$, $\mu,\,\nu=0,\dots,3$, $\tilde{m},\tilde{n}=8,9$ and $r^2=\rho^2+\left(w^8\right)^2+\left(w^9\right)^2$. Moreover, $d\Omega_3^2$ denotes the metric for the $S^3$ sphere. The D$7$-brane is embedded in such a way that $w^8$ and $w^9$ are its transverse coordinates.
The AdS radius $R$ is given in terms of the number of the background D$3$-branes, the string coupling $g_s$ and the string tension $\alpha'$ by
\begin{align}
R^4 =   4 \pi g_s N (\alpha')^2 \, .
\end{align}

Let us first consider the supergravity solution associated with
the D$3$-brane background \eqref{eq: D3 background}. It comes with the dilaton $\phi$ and a Ramond-Ramond (R-R) four-potential $C_{(4)}$ given by
\begin{equation}
\label{eq: D3 dilaton C4}
e^{\phi} = {\rm constant},\quad C_{(4)} = \frac{r^{4}}{R^4} ~ dx^0\wedge\cdots\wedge dx^4\,.
\end{equation}
This leads to the R-R five-form
\begin{equation}
\label{eq: F5}
\begin{split}
	F_{(5)}
	=
	(1+\star)dC_{(4)}
	=
	&
	~
	\frac{4}{R^4} r^2 ~ dx^0\wedge\cdots\wedge dx^3\wedge\left(\rho d\rho + w^{\tilde{m}} dw^{\tilde m}\right)
	\\
	&	
	+
	\frac{4 R^4 \rho^3 }{r^6}\left(\rho ~ \omega_{S^3}\wedge dw^{8}\wedge dw^9
		+d\rho\wedge\omega_{S^3}\wedge\left(w^8dw^9-w^9d w^8\right)		
		\right)\,,
\end{split}
\end{equation} 
where $\omega_{S^3}$ is the standard volume form of $S^3$.

In the next sections we will couple a spinor $\Psi$ to the above supergravity background. 
This requires us to introduce a local Lorentz frame that allows us to treat the metric \eqref{eq: D3 background} as locally flat. For the local Lorentz frame we introduce vielbeine
\begin{equation}
	e^I
	=
	e^I_M dx^M\,,
\end{equation}
where $I, M=0,\dots, 9$ and $I$ denotes the locally flat coordinates.
In terms of the $e^I$, the metric \eqref{eq: D3 background} is given by
\begin{equation}
	ds^2
	=
	\eta_{IJ}e^I e^J\,.
\end{equation}
For the geometry \eqref{eq: D3 background}, we obtain the following
components $e^I_M$ for the vielbeine,
\begin{align}
\label{eq: D3D7 Lorentz frame}
\begin{aligned}
e^{I}_{\mu} &= \frac{r}{R} ~ \delta^{I}_{\mu}\,, &&& e^{I}_{\rho} &= \frac{R}{r} ~ \delta^{I}_{\rho}\,,  &&&
e^{I}_{i} &= \frac{R ~ \rho}{r} ~ \hat{e}^I_i\,, &&& e^{I}_{\tilde{m}} &= \frac{R}{r} ~ \delta^{I}_{\tilde{m}}\,.
\end{aligned}
\end{align}
Here the index $i=5,6,7$ refers to the coordinates of the $S^3$ sphere in \eqref{eq: D3 background}. For $I$ referring to a coordinate on $S^3$, i.e. $I=5,6,7$, the object $\hat{e}^I_i$ is the dreibein on $S^3$. When $I$ denotes a coordinate transverse to $S^3$, i.e. $x^\mu\,, \rho$ or $w^{\tilde{m}}$, we set $\hat{e}_i^I$ to zero. 
\\

In terms of the local Lorentz frame \eqref{eq: D3D7 Lorentz frame} the five-form \eqref{eq: F5} is given by
\begin{equation}
\label{d3fiveform}
\begin{split}
 F_{(5)}=\frac{4}{R r}\Big(&e^0\wedge\cdots\wedge e^3\wedge \big(\rho~ e^\rho+w^8 e^8+w^9 e^9\big)
	\\
	&
	+
	\rho~ e^{S^3}\wedge e^8\wedge e^9
	+e^\rho\wedge e^{S^3}\wedge (w^8 e^9-w^9 e^8)
	\Big)\,,
 \end{split}
 \end{equation}
 where $e^{S^3}={e}^5\wedge {e}^6 \wedge {e}^7$ is the $\wedge$-product of the three vielbeine corresponding to the $S^3$ directions of \eqref{eq: D3 background}.
\\
 
In order to write down an action describing the dynamics of the spinor $\Psi$,
we need to compute the spin connection 
\begin{align}
	\omega^{I J}_{M}
	=
	-\omega^{JI}_M
	=
	e^I_N G^{NP}\nabla_M e^J_P
	=
	e^I_N G^{NP}\left(\partial_M e^J_P-\Gamma^R_{MP}e^J_R\right)\,.
\end{align}
The non-vanishing spin-connection components corresponding to the Lorentz frame \eqref{eq: D3D7 Lorentz frame} are given by
\begin{equation}
\label{spinconnectiond3}
\begin{split}
	\omega_\mu^{IJ}
	&
	=
	\frac{1}{R^2} \left( \rho\big(
			\delta^I_\mu\delta^J_\rho
			-
			\delta^J_\mu\delta^I_\rho
		\big)
		+
		w^{\tilde{m}}
		\big(
			\delta^I_\mu\delta^J_{\tilde{m}}
			-
			\delta^J_\mu\delta^I_{\tilde{m}}
		\big) \right)
	\,,
	\\
	\omega^{IJ}_\rho
	&
	=
	\frac{w^{\tilde{m}}}{r^2}
	\big(
		\delta^I_{\tilde{m}}\delta^J_\rho
		-
		\delta^J_{\tilde{m}}\delta^I_\rho
	\big)
	\,,
	\\
	\omega^{IJ}_{i}
	&
	=
	\hat{\omega}^{IJ}_{i}
	+
	\Big(
		\frac{\rho^2}{r^2}
		-
		1
	\Big)
	\big(
		\delta^I_\rho\hat{e}^J_{i}
		-
		\delta^J_\rho\hat{e}^I_{i}
	\big)
	+
	\frac{\rho w^{\tilde{m}}}{r^2}
	\big(
		\delta^I_{\tilde{m}}\hat{e}^J_{i}
		-
		\delta^J_{\tilde{m}}\hat{e}^I_{i}
	\big)
	\,,
	\\
	\omega^{IJ}_{\tilde{m}}
	&
	=
	\frac{\rho}{r^2}
	\big(
		\delta^I_\rho\delta^J_{\tilde{m}}
		-
		\delta^J_\rho\delta^I_{\tilde{m}}
	\big)
	+
	\frac{w^{\tilde{n}}}{r^2}
	\big(
		\delta^I_{\tilde{n}}\delta^J_{\tilde{m}}
		-
		\delta^J_{\tilde{n}}\delta^I_{\tilde{m}}
	\big)
	\,,
\end{split}
\end{equation}
where $\hat{\omega}^{IJ}_i$ is defined as the spin-connection on the three sphere
if $I$ and $J$ correspond to the $S^3$ coordinates and set to zero otherwise. 

\subsection{Embedding of the probe D7-brane}
\label{sec: D7 embedding}
We use the standard D$7$-probe brane embedding as reviewed in
\cite{Erdmenger:2007cm}. Let us briefly summarize the main features.
The D$7$-probe brane is embedded  into the D$3$-brane geometry of \Sec \ref{dbrnssugrabckgrnd}.
This embedding as well as its bosonic fields are described by 
\begin{equation}
\label{eq: SD7}
	S_{D7}
	=
	S_{DBI}
	+
	S_{WZ}\,,
\end{equation}
where $S_{DBI}$ is the Dirac-Born-Infeld (DBI) action for the D7-brane
probe, with tension $T_7$,
\begin{equation}
\label{eq: SDBI D7}
	 S_{DBI}
	 =
	 -T_7\int\! d^{8}\xi e^{-\phi}\sqrt{-\det\big(g_{AB}+2\pi\alpha'F_{AB}\big)}
\end{equation}
and the Wess-Zumino (WS) term is
\begin{equation}
	S_{WZ}
	=
	\frac{(2\pi\alpha')^2}{2}T_7\int P[C_{(4)}]\wedge F\wedge F\,.
\end{equation}
Here, $g_{AB}$ is the pullback of the metric \eqref{eq: D3 background} to the world-volume of the D$7$-brane and $F=dA$ is the field strength of the gauge field $A$ on the brane. Moreover, $P[C_{(4)}]$ is the pullback of the R-R four-potential
\eqref{eq: D3 dilaton C4} to the D$7$-brane.
We choose the $x^\mu$, $\rho$ and $S^3$ directions of \eqref{eq: D3 background} as world-volume coordinates $\xi^A$, as shown in Table \ref{tab: D7 embedding}.

\begin{table}[h]
\begin{center}
\begin{tabular}{ |c|c|c|c|c|c|c|c|c|c|c| } 
\hline
  	{coordinates} & \multicolumn{4}{c|}{$x^\mu$} & $\rho$  & \multicolumn{3}{c|}{$S^3$} &  \multicolumn{2}{c|}{$w^{\tilde{m}}$}\\
 \hline
  	{dim} & 0 & 1 & 2 & 3 & 4 & 5 & 6 & 7 &  8 & 9\\ 
  \hline
 D3 & $\times$ & $\times$ & $\times$ & $\times$ & $\cdot$ &  $\cdot$ &  $\cdot$ &  $\cdot$ &  $\cdot$ &  $\cdot$   \\ 
 \hline
 D7 & $\times$ & $\times$ & $\times$ & $\times$ & $\times$ & $\times$ & $\times$ & $\times$ &  $\cdot$ &  $\cdot$\\ 
 \hline
\end{tabular}
\end{center}
\caption{The embedding of the D$3$- and D$7$-branes. We choose the D$3$-branes to
be embedded along the directions $0,\dots,3$. We refer to these coordinates as $x^\mu$. The D$7$-brane is embedded along $0,\dots,4,5,6,7$. As can be seen from \eqref{eq: D3 background}, the 4-direction is radial and we refer to it as $\rho$. Moreover, the directions $5,6,7$ form a three-sphere \eqref{eq: D3 background}. We refer to them as $S^3$.}
\label{tab: D7 embedding}
\end{table}

The ground state embedding of the brane may be found by setting $F$ equal to zero in \eqref{eq: SD7} and using the ansatz $w^8=0$, $w^9=L(\rho)$. This leads to
\begin{equation}
	S_{D7}
	=
	-T_{7} \operatorname{vol}\left(S^3\right) \int\! d^4 x ~ d \rho~ e^{-\phi}~ \rho^3 \sqrt{1+ (\partial_\rho L)^2}\,.
\end{equation}
Since $e^{-\phi}$ is constant (see \eqref{eq: D3 dilaton C4}), it is easy to verify that this action is minimized if $L$ is constant. Thus the brane wraps the $x^\mu$, $\rho$ and $S^3$ directions at a constant value $L$ of $w^9$.
This leads to the following metric on the brane,
\begin{align}
\label{eq: D7 metric}
	ds^2_{D7}
	=
	g_{AB}d\xi^Ad\xi^B
	=
	\frac{r^{2}}{R^2} \eta_{\mu\nu}dx^\mu dx^\nu + \frac{R^{2}}{r^2} d \rho^2  + \frac{R^2 \rho^2}{r^2} d \Omega_{3}^2\,,
\end{align}
where $r^2=\rho^2+L^2$.
This flat embedding of the D$7$-brane preserves half of the original (sixteen) supercharges of the D$3$-brane background. 
The distance $L$ between the D$7$- and D$3$-branes corresponds to the quark mass $m_q$. We have $L=2\pi\alpha' m_q$.
$L$ sets the mass scale of the theory and its bound states. 

\subsection{Bosonic fluctuations}
\label{sec: D3D7 bosonic fluctuations}

The bosonic fluctuations of the embedding of the D$7$-brane and the
gauge field are studied in  \cite{Kruczenski:2003be}. These fluctuations correspond to scalar and vector mesons
on the boundary.
In this section we briefly review the results of \cite{Kruczenski:2003be}.

\subsubsection{Scalar fluctuations}
\label{sec: scalar fluctuations}
We consider small fluctuations $\Phi(\xi^A)$ of the D$7$ brane transverse to the flat embedding $w^9=L$ presented in \Sec \ref{sec: D7 embedding}. These correspond to scalar mesons on the boundary.
By using\footnote{Both fluctuations in \eqref{eq: scalar fluctuation} provide the same equation of motion for $\Phi$.}
\begin{equation}
\label{eq: scalar fluctuation}
	w^8
	=
	2\pi\alpha'\Phi
	\quad\text{or}\quad
	w^9
	=
	L+2\pi\alpha'\Phi, 
\end{equation}
in the DBI action \eqref{eq: SDBI D7} for $F_{AB}=0$
and expanding to leading order in $\Phi$, we obtain a second order partial differential equation for $\Phi$. To solve this equation, we make the plane wave ansatz
$\Phi=f(\rho) e^{i k^{\mu} x_{\mu}} \mathcal{Y}^{\ell}$, where $\mathcal{Y}^{\ell}$ is a scalar spherical harmonic on $S^3$ satisfying
\begin{equation}
	\nabla^2 \mathcal{Y}^{\ell} = -\ell(\ell+2) \mathcal{Y}^{\ell},
\end{equation}
for $\ell\in \mathbb{N}_0$.
This leads to
\begin{align}
	\partial_{\rho}^2 f(\rho)
	+
	\frac{3}{\rho} \partial_{\rho} f(\rho)
	+
	\frac{R^4 ~ M^2}{(\rho^2+L^2)^2} f(\rho)
	-
	\frac{\ell(\ell+2)}{\rho^2} f(\rho)
	=
	0
	\,, \label{scalareomd3d7}
\end{align}
where $M^2=-k^2$ corresponds to the mass of the mesons dual to $\Phi$.
Solving \eqref{scalareomd3d7} and imposing normalizability we find the mass spectrum to be of the discrete form
\begin{align}
M_s = 2 \frac{L}{R^2} \sqrt{(n+\ell+1)(n+\ell+2)}\,,\quad\text{where}\quad n \in \mathbb{N}_0 \quad\text{and}\quad\ell \in \mathbb{N}_0\,.
\label{scalarspectrum}
\end{align}
In particular, we see that the meson mass scales with $L$, i.e. the distance between the probe D$7$-brane and the stack of D$3$-branes and is thus proportional to the quark mass. The solution of \eqref{scalareomd3d7} corresponding to the mass \eqref{scalarspectrum} is given by
\begin{align}
f(\rho) &= \frac{\rho^{\ell}}{(\rho^2+L^2)^{n+\ell+1}} {_2}F_1 \Big(-(n+\ell+1),-n,\ell+2,-\frac{\rho^2}{L^2}\Big)\,.
\label{scalarmodes}
\end{align}

The solution above has the near-boundary $\left( \rho \rightarrow \infty \right)$ expansion 
$f(\rho) \sim 1/\rho^{\ell+2}$ 
and the conformal dimension of the dual operator is 
$\Delta = \ell+3$.

\subsubsection{Vector fluctuations}
\label{sec: vector fluctuations}
In analogy to the scalar fluctuations $\Phi$, small excitations of the gauge field
$A$ appearing in \eqref{eq: SDBI D7} may be considered. They correspond to vector mesons. By imposing the gauge fixing condition $\partial_\mu A^\mu=0$, three types of gauge fields may be distinguished \cite{Kruczenski:2003be},
\begin{align}
\label{eq: type I}
&
\text{Type I} &&  A_{\mu}= 0\,, \quad  A_{\rho} = 0\,, \quad A_i = h^{\pm}(\rho) e^{i k \cdot x} \mathcal{Y}^{\ell,\pm}_i\,,
\\
\label{eq: type II}
&
\text{Type II} &&  A_{\mu} = \zeta_{\mu} ~ g(\rho) e^{i k \cdot x} \mathcal{Y}^{\ell}\,, \quad A_{\rho} = 0\,, \quad A_i = 0\,,
\\
\label{eq: type III}
&
\text{Type III} &&  A_{\mu} =0\,, \quad A_{\rho} = y(\rho) e^{i k \cdot x} \mathcal{Y}^{\ell}\,, \quad A_i =  \tilde{y}(\rho) e^{i k \cdot x} \nabla_i \mathcal{Y}^{\ell}\,.
\end{align}
Here $\mathcal{Y}^{\ell,\pm}_i$ and $\mathcal{Y}^{\ell}$ are spherical harmonics on $S^3$ and $\zeta_\mu k^\mu=0$ guarantees the gauge fixing $\partial_\mu A^\mu$ for Type II.
By expanding \eqref{eq: SDBI D7} to leading order in $A$ we obtain a second order partial differential equation for $A$, 
\begin{align}
\partial_{A} (\sqrt{-g} F^{AB}) - 4 \frac{\rho}{R^4} (\rho^2 + L^2) \delta^{B}_{i} \epsilon^{i j k} A_{i} \partial_{j} A_{k} = 0\,,
\label{gaugeeomgeneral}
\end{align}
where $i,j,k=5,6,7$ correspond to the $S^3$ directions of the brane.
By introducing the symbol $\delta^B_i$ in \eqref{gaugeeomgeneral} we emphasize that the corresponding term only vanishes if $B\neq 5,6,7$.  
The solutions for the three types of $A$ \eqref{eq: type I} -- \eqref{eq: type III} are given in Table \ref{tab: gauge boson solution D7} \cite{Kruczenski:2003be}.
\begin{table}[h]
\begin{center}
\begin{tabular}{ |c|l|c| } 
\hline
  	{Type} & \multicolumn{1}{c|}{Solution} & $\ell$\\
 \hline
  	\multirow{ 2}{*}{I} & $h^{+}(\rho) = \rho^{\ell+1} ~ {_2}F_1 \left(-(n+\ell+3), -n , \ell+2, -\rho^2/L^2\right)/(\rho^2 + L^2)^{n+\ell+3}$ & $\geq 1$\\
  \cline{2-3}
  & $h^{-}(\rho) = \rho^{\ell+1} ~ {_2}F_1 \left(-(n+\ell+1), -n , \ell+2, -\rho^2/L^2\right)/(\rho^2 + L^2)^{n+\ell+1}$ & $\geq 1$\\ 
 \hline
  II & $g(\rho) = \rho^{\ell} ~ {_2}F_1 \left(-(n+\ell+1),-n,\ell+2,-\rho^2/L^2 \right)/(\rho^2+L^2)^{n+\ell+1}$ & $\geq 0$\\ 
 \hline
 \multirow{ 2}{*}{III} & $y(\rho) = \rho^{\ell - 1} ~ {_2}F_1 \left(-(n+\ell+1),-n,\ell+2,-\rho^2/L^2\right)/(\rho^2+L^2)^{n+\ell+1}$ & \multirow{ 2}{*}{$\geq 1$}
 \\
 \cline{2-2}
 & $\tilde{y}(\rho)=\partial_{\rho}(\rho^3 y(\rho))/\rho ~ \ell(\ell+2)$ &\\
 \hline
\end{tabular}
\end{center}
\caption{The three types of gauge fields. By inserting the ans\"atze \eqref{eq: type I}, \eqref{eq: type II} and \eqref{eq: type III} for the gauge fields of type I, II and III into the equation of motion \eqref{gaugeeomgeneral}, we find the listed solutions.}
\label{tab: gauge boson solution D7}
\end{table}

The appearence of the parameter $n\in\mathbb{N}_0$ is a result of the normalizability of the solutions and leads to the discrete mass spectrum
and conformal dimensions $\Delta$ given in Table \ref{tab: gauge boson mass spectrum D7}.
\begin{table}[h]
\begin{center}
\begin{tabular}{ |c|l|c| } 
\hline
  	{Type} & \multicolumn{1}{c|}{Mass} & $\Delta$\\
 \hline
  	\multirow{ 2}{*}{I} & $M_+ = 2 ( L/R^2) \sqrt{(n+\ell+2)(n+\ell+3)}$ & $\ell +5$\\
  \cline{2-3}
  & $M_- = 2 ( L/R^2) \sqrt{(n+\ell)(n+\ell+1)}$ & $\ell +1$\\ 
 \hline
  II & $M_{II} = 2 ( L/R^2) \sqrt{(n+\ell+1)(n+\ell+2)}$ & $\ell+3$\\ 
 \hline
 III & $M_{III} = 2( L/R^2) \sqrt{(n+\ell+1)(n+\ell+2)}$ & $\ell+3$
 \\
 \hline
\end{tabular}
\end{center}
\caption{The mass spectra for the three types of gauge fields \eqref{eq: type I}, \eqref{eq: type II} and \eqref{eq: type III}. The solutions of the gauge fields of type I, II and III given in Table \ref{tab: gauge boson solution D7} come with the listed discrete mass spectra. The discreteness of the spectra is a consequence of the normalizability condition imposed to the solutions.}
\label{tab: gauge boson mass spectrum D7}
\end{table}
Just as for the scalar mesons (see \Sec \ref{sec: scalar fluctuations}) the mass of the vector mesons scales with the distance $L$ of the D$7$-brane from the stack of D$3$ branes.

\section{Fermionic fluctuations in the D3/D7 system}
\label{sec: fermionic fluctuations in D3/D7}
Our main focus are the fermionic fluctuations. These are
dual to mesinos, i.e. the fermionic superpartners of the mesons. The
fermionic excitations of D$7$-branes have been studied in
\cite{Kirsch:2006he} for the case of massless quarks, i.e. $L=0$. In
the following we provide a full derivation for the massive case,
i.e. $L>0$.

\subsection{Fermionic part of the D$7$-brane action}
The supersymmetric completion of the bosonic D7-brane action in the D3-background is given by the fermionic action \cite{Martucci:2005rb}
\begin{align}
\label{eq: Sf with sigma2}
	S^{f}_{D7}
	=
	\frac{T_{D7}}{2} \int d^8 \xi
		\sqrt{- \det g_{AB}} \bar{\Psi} \mathcal{P}_{-} \Gamma^{A} \Big(
			D_{A}
			+
			\frac{1}{2 \times 8 \times 5!} F_{N P Q R S} \Gamma^{N P Q R S}
			~\left(i \sigma_{2}
		\right) ~ \Gamma_{A}  \Big) \Psi 
\,.
\end{align}
Here, $\Psi$ is a ten-dimensional pair of positive-chirality Majorana-Weyl spinors of type IIB supergravity written in the doublet spinor notation.
The $\Gamma_A$ are the ten-dimensional $\Gamma$-matrices on AdS$_5\times S^5$ pulled back to the worldvolume of the probe D7-brane,
\begin{equation}
\label{eq: pulled back gamma}
	\Gamma_{A} = \Gamma_{M} \partial_{A} x^{M}\,.
\end{equation}
 Moreover, $\mathcal{P}_{-}$ is a $\kappa$-symmetry projector, and $D_{A}$ is the curved-spacetime gauge covariant derivative.
By considering the map $i\sigma_2\Psi=-i\Psi$ with the Pauli matrix $\sigma_2$ in \eqref{eq: Sf with sigma2} we may pass from two real spinors to one complex spinor. Thus we end up with an action of the form 
\begin{align}
\label{fermionicdbrane}
S^{f}_{D7} = \frac{T_{D7}}{2} \int d^8 \xi \sqrt{- \det g_{AB}} \bar{\Psi} \mathcal{P}_{-} \Gamma^{A} \Big(D_{A} - \frac{i}{2 \times 8 \times 5!} F_{N P Q R S} \Gamma^{N P Q R S} \Gamma_{A}  \Big) \Psi\,.
\end{align}

\subsection{\boldmath Decomposition of the $\Gamma$ matrices and spinors} \label{gammasection}

In order to solve the equation of motion derived from \eqref{fermionicdbrane}, it is useful to have an explicit basis for the Dirac matrices.
The ten-dimensional curved spacetime Dirac matrices $\Gamma^M$ are related to the
Dirac matrices on the local Lorentz frame \eqref{eq: D3D7 Lorentz frame} via
\begin{equation}
\label{eq: st gamma ito lf gamma}
	\Gamma_M
	=
	e^I_M\Gamma_I\,.
\end{equation}
We consider the following decomposition\footnote{This decomposition is a generalization of the one used in \cite{Kirsch:2006he} for the special case $L=0$.} of the flat ten-dimensional Dirac matrices $\Gamma^I$,
\begin{align}
\label{gammadecomposition}
\Gamma^{\alpha} &= \sigma_{2} \otimes \mathbf{1}_{4} \otimes \gamma^{\alpha}\,,
\\
\label{eq: gamma m}
\Gamma^{m} &= \sigma_{1} \otimes \gamma^{m} \otimes \mathbf{1}_{4}\,,
\end{align} 
where in the above $\alpha=0,1,2,3,\rho$ and $m= 5, 6, 7, 8,9$ is taking values on the $3$-sphere and in the directions transverse to the D$7$-brane (see Table \ref{tab: D7 embedding}). We use the shorthand notation $S^3=(5, 6, 7)$ in the following.
The lower-case $\gamma$ are $4\times 4$ matrices satisfying the Clifford algebras 
\begin{align}
	\{ \gamma^{\alpha}, \gamma^{\beta} \}= 2 \eta^{\alpha \beta}\,,
	\quad\quad
	\{ \gamma^m, \gamma^{n} \}= 2 \delta^{mn}\,.
\end{align}
So the $\gamma^\alpha$ and $\gamma^m$ obey a Minkowskian and Euclidean Clifford algebra, respectively.
As usual we define Pauli matrices
\begin{equation}
    \sigma_{1} =
    \begin{pmatrix}
    0 & 1 \\
    1 & 0 \\
    \end{pmatrix}
    ,\quad
    \sigma_{2} =
    \begin{pmatrix}
    0 & -i\\
    i & 0
    \end{pmatrix}
    ,\quad
    \sigma_{3} =
    \begin{pmatrix}
    1 & 0\\
    0 & -1
    \end{pmatrix}\,. 
\end{equation}
\eqref{gammadecomposition} and \eqref{eq: gamma m} provide a $32$-dimensional representation of the $(9+1)$-dimensional Minkowskian Clifford algebra, i.e.
\begin{equation}
	\{\Gamma^\alpha,\Gamma^\beta\}=2\eta^{\alpha\beta}\,,
	\quad
	\{\Gamma^m,\Gamma^n\}=2\delta^{m n}\,,
	\quad
	\{\Gamma^\alpha,\Gamma^m\}=0\,.
\end{equation}
We can define ``$\gamma^5$'' type matrices for each sub-space as the product of all $\gamma$ matrices for the subspace,
\begin{align}
\label{eq: Gamma^AdS, Gamma^S}
\Gamma^{AdS} &= \Gamma^{0123\rho}= i \sigma_{2} \otimes \mathbf{1}_{4} \otimes \mathbf{1}_{4}\,, &&& \Gamma^{S} &=\Gamma^{S^3 89} = \sigma_{1} \otimes  \mathbf{1}_{4} \otimes \mathbf{1}_{4}\,.
\end{align}
The following raising and lowering relations apply,
\begin{align}
\Gamma^{AdS} &= - \Gamma_{AdS}  &&&   \Gamma^{S} &= \Gamma_{S}\,.
\end{align}
Moreover, using the above relations the chirality operator $\Gamma^{11}=\Gamma^{AdS}\Gamma^S$ can be written as 
\begin{align}
\Gamma^{11} = \sigma_{3}  \otimes \mathbf{1}_{4}  \otimes \mathbf{1}_{4}\,.
\end{align}
The ten-dimensional spinor $\Psi$ has positive chirality, $\Gamma^{11} \Psi = \Psi$. We choose the following decomposition
\begin{equation}
\label{eq: spinor decomposition D7}
\Psi = \uparrow \otimes \chi \otimes \hat{\psi}\,,\quad\text{where}\quad\uparrow = \begin{pmatrix}
           1 \\
           0 \\
         \end{pmatrix},
\end{equation}
and the spinors $\chi$ and $\hat{\psi}$ both have four entries. This decomposition matches the one of the Dirac matrices \eqref{gammadecomposition}, \eqref{eq: gamma m}, i.e. the Pauli matrix part of the Dirac matrices acts on $\uparrow$ while the first $4\times 4$ part acts on $\chi$ and the second on $\hat{\psi}$.

\subsection{Dirac equation}
The action \eqref{fermionicdbrane} leads to the Dirac equation
\begin{equation}
\label{eq: eom D7 spinor}
	\slashed{D}\Psi - \frac{i}{1920} \Gamma^A F_{N P Q R S} \Gamma^{N P Q R S} \Gamma_{A}\Psi
	=
	0,
\end{equation}
on the D$7$-brane, where $\slashed{D}=\Gamma^AD_A$.
We now aim at reducing \eqref{eq: eom D7 spinor} to an equation for the
spinor $\hat{\psi}$ appearing in the decomposition \eqref{eq: spinor decomposition D7} of $\Psi$. For this we make the ansatz
\begin{equation}
\label{eq: chi and psi dependence}
	\chi
	=
	\chi(S^3)
	\quad\text{and}\quad
	\hat{\psi}
	=
	\hat{\psi}(x^\mu,\rho)\,,
\end{equation}
i.e. we assume the dependence of $(x^\mu,\rho)$ and the $S^3$ coordinates to factorize in $\Psi$.
The covariant derivative $D_A$ in the kinetic term of \eqref{eq: eom D7 spinor}
corresponds to the extrinsic curvature of the D$7$-brane and thus is given by\footnote{We thank L. Martucci and D. Van den Bleeken for clarifying this.}
\begin{equation}
	D_A
	=
	\partial_A
	+
	\frac{1}{8}\partial_Ax^M\omega_M^{IJ}[\Gamma_I,\Gamma_J]\,.
\end{equation} 
Therefore, the kinetic term in \eqref{eq: eom D7 spinor} is given by
\begin{equation}
	\slashed{D}\Psi
	=
	g^{AB}\partial_A x^M\Gamma_M D_B\Psi
	=
	g^{AB}\partial_A x^M\Gamma_M \left(
		\partial_B
		+
		\frac{1}{8}\partial_Bx^M\omega_M^{IJ}[\Gamma_I,\Gamma_J]
	\right)\Psi\,,
\end{equation}
where we have used \eqref{eq: pulled back gamma}. Inserting \eqref{spinconnectiond3}, \eqref{eq: D7 metric}
as well as applying $\{\Gamma_I,\Gamma_J\}=2\eta_{IJ}$ leads to
\begin{equation}
\label{eq: slashed D in D7}
	\slashed{D}\Psi
	=
	\left(\frac{R}{r} ~ \Gamma^{\mu} ~ \partial_{\mu} + \frac{r}{R} ~  \Gamma^{\rho} ~ \partial_{\rho} + \frac{r}{R ~ \rho} \slashed{\nabla}_{S^{3}} + \frac{1}{2 R} \left(\frac{\rho}{r} +  3 \frac{r}{\rho} \right) \Gamma^{\rho}\right)\Psi
	\,,
\end{equation}	
where the Dirac matrices belong to the Lorentz frame and
\begin{equation}
	 \slashed{\nabla}_{S^{3}}
	 =
	 g_{S^3}^{ij}\hat{e}^k_i\Gamma_k\nabla_{S^3j},
\end{equation}
is the covariant derivative on the three sphere corresponding to the directions $5,6,7$ (see \Sec \ref{sec: D7 embedding}). In the above $g_{S^3}$ is the $S^3$ metric and $\Gamma_k$ are the Lorentz frame Dirac matrices in directions $5,6,7$. 

In \eqref{eq: slashed D in D7} we see that neither $\Gamma^8$ nor
$\Gamma^9$ appears in $\slashed{D}$. This is a consequence of the flat
embedding of the D$7$-brane along $x^\mu$, $\rho$ and $S^3$ at a
constant value $w^9=L$. In the more general situation of non-flat
embeddings for which the distance of the brane in $w^9$ direction
depends on the value of $\rho$, contributions of $\Gamma^9$ are to be
expected. We leave this case to future work. 

By considering the decompositions \eqref{gammadecomposition}, \eqref{eq: gamma m} and \eqref{eq: spinor decomposition D7} of the Dirac matrices and the spinor together with \eqref{eq: chi and psi dependence}, we find
\begin{equation} \label{eq: dirac_operator_need_a_label}
\begin{split}
	\slashed{D}\Psi
	=
	&
	\downarrow\otimes\chi\otimes i\left(
		\frac{R}{r} ~ \gamma^{\mu} ~ \partial_{\mu}
		+
		\frac{r}{R} ~  \gamma^{\rho} ~ \partial_{\rho}
		+
		\frac{1}{2 R} \left(\frac{\rho}{r}
		+
		3 \frac{r}{\rho} \right) \gamma^{\rho}		
		\right)\hat{\psi}
	\\
	&
	+
	\downarrow\otimes\frac{r}{R ~ \rho} \slashed{\nabla}_{S^{3}}\chi\otimes\hat{\psi}
	\,.
\end{split}
\end{equation}
Here, on the r.h.s~of the equation the Feynman slash refers to
contractions 
with the lower case $\gamma$
matrices appearing in the decomposition \eqref{gammadecomposition},
\eqref{eq: gamma m} of $\Gamma^I$. The $\gamma^{\rho}$ is the chiral
$\gamma$-matrix in the bulk flat directions that upon acting on the
spinor $\hat{\psi}$ will give two eigenvalues $\pm 1$.  In a
two-component matrix notation, $\gamma^\rho$ can be expressed as the diagonal matrix with $\pm 1$ as its entries $\gamma^{\rho} =$diag$(1,-1)$. 
We now make the ansatz of $\chi$ being a spinor spherical harmonic, i.e. $\chi=\chi^\pm_\ell$, with \cite{Camporesi:1995fb}
\begin{equation}
\label{eq: chi spherical harmonic}
	\slashed{\nabla}_{S^3}\chi^\pm_\ell
	=
	\pm i\left(\ell
		+
		\frac{3}{2}
		\right)\chi^\pm_\ell\,.
\end{equation}
This results in
\begin{equation}
\label{eq: kinetic part of dirac eq}
	\slashed{D}\Psi
	=
	\downarrow\otimes\chi^\pm_\ell\otimes i\left(
		\frac{R}{r} ~ \gamma^{\mu} ~ \partial_{\mu}
		+
		\frac{r}{R} ~  \gamma^{\rho} ~ \partial_{\rho}
		+
		\frac{1}{2 ~ R} \left(\frac{\rho}{r}
		+
		3 \frac{r}{\rho}
		\right) \gamma^{\rho}
		\pm
		\frac{r}{R ~ \rho}
		\left(\ell
		+
		\frac{3}{2}
		\right)	
		\right)\hat{\psi}
	\,.
\end{equation}
So by choosing $\chi$ to be a spinor spherical harmonic on $S^3$, $\slashed{D}$ may be
formulated as an operator that only acts on $\hat{\psi}(x^\mu,\rho)$. An analogous result can be derived for the second term in \eqref{eq: eom D7 spinor},
\begin{equation}
\label{eq: RR part D7}
	\frac{i}{1920} \Gamma^A F_{N P Q R S} \Gamma^{N P Q R S} \Gamma_{A}\Psi\,,
\end{equation}
as we now show.
When using the local Lorentz frame for the R-R five-form \eqref{d3fiveform},
we may work with the frame Dirac matrices $\Gamma^I$ instead of the curved spacetime $\Gamma^M$, i.e.
\begin{equation}
	\frac{i}{1920} \Gamma^A F_{N P Q R S} \Gamma^{N P Q R S} \Gamma_{A}\Psi
	=
	\frac{i}{1920} \Gamma^A F_{I_1 I_2 I_3 I_4 I_5} \Gamma^{I_1 I_2 I_3 I_4 I_5} \Gamma_{A}\Psi\,.
\end{equation}
Here $F_{I_1 I_2 I_3 I_4 I_5}$ are the components of $F_{(5)}$ according to
the Lorentz frame,
\begin{equation}
	F_{(5)}
	=
	\frac{1}{5!}F_{I_1 \cdots I_5}e^{I_1}\wedge\cdots\wedge e^{I_5}\,.
\end{equation}
By expressing the Dirac matrices $\Gamma_A$ pulled back to the $(7+1)$-dimensional world-volume of the D$7$-brane in terms of the ten-dimensional Lorentz frame Dirac matrices $\Gamma_I$ via 
\eqref{eq: D3D7 Lorentz frame}, \eqref{eq: pulled back gamma} and \eqref{eq: st gamma ito lf gamma}
and applying $\{\Gamma_I,\Gamma_J\}=2\eta_{IJ}$, we obtain
\begin{equation}
\label{eq: R-R term short D7}
	\frac{i}{1920} \Gamma^A F_{I_1 I_2 I_3 I_4 I_5} \Gamma^{I_1 I_2 I_3 I_4 I_5} \Gamma_{A}\Psi
	=
	i~\frac{\rho}{R ~ r}\Gamma^{0123\rho}\Psi
	\,.
\end{equation}
Note that in the derivation of \eqref{eq: R-R term short D7} we have used the positive chirality of $\Psi$, i.e.
\begin{equation}
	\Gamma^{11}\Psi=\Gamma^{0123\rho S^3 89}\Psi=\Psi
	\quad\Leftrightarrow\quad
	\Gamma^{0123\rho}\Psi
	=
	-\Gamma^{S^3 89}\Psi\,,
\end{equation}
Inserting the Dirac matrix and spinor decompositions \eqref{gammadecomposition}, \eqref{eq: gamma m} and \eqref{eq: spinor decomposition D7}
we find by considering \eqref{eq: Gamma^AdS, Gamma^S},
\begin{equation}
\label{eq: R-R part of dirac eq}
	\frac{i}{1920} \Gamma^A F_{I_1 I_2 I_3 I_4 I_5} \Gamma^{I_1 I_2 I_3 I_4 I_5} \Gamma_{A}\Psi
	=
	\downarrow\otimes\chi\otimes \left(-i~\frac{\rho}{R ~ r}\right)\hat{\psi}
	\,.
\end{equation}
So just as for the kinetic term $\slashed{D}\Psi$, we can express the second term
of the Dirac equation \eqref{eq: eom D7 spinor} as an operator solely acting on $\hat{\psi}$.
Combining \eqref{eq: kinetic part of dirac eq} and \eqref{eq: R-R part of dirac eq} we may write
\begin{equation}
\begin{split}
	&
	\slashed{D}\Psi - \frac{i}{1920} \Gamma^A F_{N P Q R S} \Gamma^{N P Q R S} \Gamma_{A}\Psi
	\\
	=
	&
	\downarrow\otimes\chi_\ell^\pm\otimes i\left(
		\frac{R}{r} ~ \gamma^{\mu} ~ \partial_{\mu}
		+
		\frac{r}{R} ~  \gamma^{\rho} ~ \partial_{\rho}
		+
		\frac{1}{2 R} \left(\frac{\rho}{r}
		+
		3 \frac{r}{ \rho}
		\right) \gamma^{\rho}
		+	
		\frac{\rho}{R ~ r}
		\pm
		\frac{r}{R ~ \rho}
		\left(\ell
		+
		\frac{3}{2}
		\right)
	\right)\hat{\psi}
	=
	0
	\,.
\end{split}
\end{equation}
So we may formulate the Dirac equation \eqref{eq: eom D7 spinor} on the D$7$-brane as an equation for $\hat{\psi}$, given by
\begin{equation}
\label{eq: 1st order dirac eq}
	\left(\frac{R}{r} ~ \gamma^{\mu} ~ \partial_{\mu}
			+
			\frac{r}{R} ~  \gamma^{\rho} ~ \partial_{\rho}
			+
			\frac{1}{2 R} \left(\frac{\rho}{r}
			+
			3 \frac{r}{\rho}
			\right) \gamma^{\rho}
			+	
			\frac{\rho}{R ~ r}
			\pm
			\frac{r}{R ~ \rho}
			\left(\ell
			+
			\frac{3}{2}
			\right)
		\right)\hat{\psi}(x^\mu,\rho)
	=
	0		
	\,,
\end{equation}
The $\pm$ in front of the last term gives rise to two different sets of modes. We will refer to the operators dual to these modes as $\mathcal{G}$ (for the $+$) and $\mathcal{F}$ (for the $-$). 

A good consistency check of our results is to take the conformal limit $L \rightarrow 0$. Upon taking this limit, the world-volume geometry of the D$7$ brane returns to being AdS$_5 \times$ S$^3$ and we should reproduce the results of previous works \cite{Kirsch:2006he}, \cite{Ammon:2010pg}. 

When $L=0$, $r=\rho$ and the Dirac equations for the $\mathcal{G}$ and $\mathcal{F}$ modes reduce to Dirac equations on AdS$_5$, 
\begin{align}
\label{eq: G F eom L=0}
	\left(\slashed{D}_{AdS}
		+
		\ell
		+\frac{5}{2}
	\right)\hat{\psi}_{\mathcal{G}}^\ell
	=
	0		
	\,,\quad
	\left(\slashed{D}_{AdS}
			-
			\left(
			\ell
			+\frac{1}{2}\right)
		\right)\hat{\psi}_{\mathcal{F}}^\ell
	=
	0	
	\,,
\end{align}
where
\begin{equation}
	\slashed{D}_{AdS}
	=
	\frac{R}{\rho} ~ \gamma^{\mu} ~ \partial_{\mu}
	+
	\frac{\rho}{R} ~  \gamma^{\rho} ~ \partial_{\rho}
	+
	\frac{2}{R} \gamma^{\rho},
\end{equation}
is the covariant derivative on AdS$_5$. The AdS bulk masses $m_{\mathcal{G}}$ and $m_{\mathcal{F}}$ of $\hat{\psi}^\ell_{\mathcal{G},\mathcal{F}}$ satisfy
\begin{equation}
	|m_{\mathcal{G}}|
	=
	\ell+\frac{5}{2}\,,
	\quad
	|m_{\mathcal{F}}|
	=
	\ell+\frac{1}{2}\,,
\end{equation}
which is in agreement with \cite{Kirsch:2006he}, \cite{Ammon:2010pg}. 

Note that the Ramond-Ramond/spinor coupling \eqref{eq: RR part D7}
induced the term $\rho/r$ in \eqref{eq: 1st order dirac eq}. In the
conformal limit, this corresponds to a shift in the bulk fermion mass
by one unit, an observation first made in the dilatino spectrum of type IIB supergravity compactified on a five-sphere \cite{Kim:1985ez}.

\subsection{Second-order equations of motion}
\label{sec: 2nd order eom D7 derivation}
In order to proceed and determine the mass spectrum of the mesinos for
the $\mathcal{G}$ and $\mathcal{F}$ modes we now  construct a second
order differential equation for
$\hat{\psi}_{\mathcal{F},\mathcal{G}}^\ell$. We begin by considering the plane-wave ansatz
\begin{equation}
\label{eq: plane wave spinor}
	\hat{\psi}^\ell_{\mathcal{F},\mathcal{G}}(x^\mu,\rho)
	=
	e^{i k_\mu x^\mu} \left(  \psi^\ell_{\mathcal{F},\mathcal{G},+}(\rho) \alpha_{+} + \psi^\ell_{\mathcal{F},\mathcal{G},-}(\rho) \alpha_{-}  \right) ,
\end{equation}
where the $\alpha_{\pm}$ are eigenstates of the $\gamma^{\rho}$ satisfying $\gamma^{\rho} \alpha_{\pm} = \pm \alpha_{\pm}$, and these eigenspinors are related via 
\begin{align}
\label{eq: chiral eigenspinors}
\alpha_{-} = \frac{i ~ k_{\mu} ~ \gamma^{\mu}}{M} \alpha_{+} \, .
\end{align} 
Note that the relation \eqref{eq: chiral eigenspinors} was already used in \cite{Mueck:1998iz} for
spinors in AdS$_{d+1}$. 
The normalization is chosen such that $\alpha_{-}^\dagger \alpha_{-}=1$ - in fact though our choice only works in the rest frame of the mesino where $k^\mu=(M,0,0,0)$ (here $\gamma^{0 \dagger} = - \gamma^0$), but this choice of frame is sufficient to determine the spectrum.
We will first restrict to the case of the $\mathcal{G}$-modes here as an illustrative example. Inserting eq.($\ref{eq: plane wave spinor}$) and ($\ref{eq: chiral eigenspinors}$) in eq.($\ref{eq: 1st order dirac eq}$) leads to 
\begin{align}
\begin{aligned}
&\left( \frac{r}{R} \partial_{\rho} \psi^{\ell}_{\mathcal{G},+}(\rho) + (A+B) \psi^{\ell}_{\mathcal{G},+}(\rho) + \frac{R M}{r} \psi^{\ell}_{\mathcal{G},-}(\rho) \right) \alpha_{+} &= 0 , \\ 
&\left( - \frac{r}{R} \partial_{\rho} \psi^{\ell}_{\mathcal{G},-}(\rho) - (A-B) \psi^{\ell}_{\mathcal{G},-}(\rho) + \frac{R M}{r} \psi^{\ell}_{\mathcal{G},+}(\rho)  \right) \alpha_{-} &= 0,
\end{aligned}
\end{align}
where the factors $A$, and $B$ are given by 
\begin{align}
\begin{aligned}
\label{eq: A and B factors}
A &= \frac{1}{2 R} \left( \frac{\rho}{r} + 3 \frac{r}{\rho}  \right), &&& B &= \frac{\rho}{R r} + \frac{r}{R \rho} \left( \ell + \frac{3}{2} \right).
\end{aligned}
\end{align}
Since the spinors $\alpha_{\pm}$ are linearly independent, we conclude that
\begin{align}
\label{eq: coupled system}
\begin{aligned}
 \frac{r}{R} \partial_{\rho} \psi^{\ell}_{\mathcal{G},+}(\rho) + (A+B)
 \psi^{\ell}_{\mathcal{G},+}(\rho) + \frac{R M}{r}
 \psi^{\ell}_{\mathcal{G},-}(\rho) &= 0 \, , \\
  - \frac{r}{R} \partial_{\rho} \psi^{\ell}_{\mathcal{G},-}(\rho) -
  (A-B) \psi^{\ell}_{\mathcal{G},-}(\rho) + \frac{R M}{r}
  \psi^{\ell}_{\mathcal{G},+}(\rho) &= 0 \, .
\end{aligned}
\end{align}
From this set of coupled differential equations, we rearrange the second to obtain 
\begin{align}
\label{eq: first order sltn}
\psi^{\ell}_{\mathcal{G},+}(\rho) = \frac{r}{R M} \left(
  \frac{r}{R} \partial_{\rho} \psi^{\ell}_{\mathcal{G},-}(\rho) +
  (A-B) \psi^{\ell}_{\mathcal{G},-}(\rho) \right) \, .
\end{align}
If we now insert  ($\ref{eq: first order sltn}$) in the first equation of the coupled system eq.($\ref{eq: coupled system}$) we obtain the desired second order equation
\begin{align}
\label{eq: second order psi minus}
\left( \frac{r^2}{R^2} \partial_{\rho}^2 + \frac{3}{R^2} \left( \rho + \frac{r^2}{\rho} \right) \partial_{\rho} + \frac{M^2 R^2}{r^2} - \frac{r^2 \left(\ell^2 + 2 \ell \right)}{R^2 \rho^2} - \frac{4 \ell +2}{R^2} -\frac{3 \rho^2}{4 R^2 r^2} \right) \psi^{\ell}_{\mathcal{G},-}(\rho) = 0.
\end{align}
Thus we need to solve ($\ref{eq: second order psi minus}$) for
$\psi^{\ell}_{\mathcal{G},-}(\rho)$. Subsequently, we may insert the
solution into ($\ref{eq: first order sltn}$) to obtain $\psi^{\ell}_{\mathcal{G},+}(\rho)$. 

In a complementary approach, we are able to obtain the equivalent second-order equations of motion for the $\psi^{\ell}_{\mathcal{G},+}$. We proceed as follows: From ($\ref{eq: coupled system}$) we solve the first one to obtain 
\begin{align}
\label{eq: first order sltn second}
\psi^{\ell}_{\mathcal{G},-}(\rho) = \frac{r}{R M} \left(-  \frac{r}{R} \partial_{\rho} \psi^{\ell}_{\mathcal{G},+}(\rho) - (A+B) \psi^{\ell}_{\mathcal{G},+}(\rho) \right),
\end{align}
and as before we insert this solution in the second equation from the set of the first-order coupled ones, eq.($\ref{eq: coupled system}$) and obtain 
\begin{align}
\label{eq: second order psi plus}
\left( \frac{r^2}{R^2} \partial_{\rho}^2 + \frac{3}{R^2} \left( \rho + \frac{r^2}{\rho} \right) \partial_{\rho} + \frac{M^2 R^2}{r^2} - \frac{r^2 \left(\ell^2 + 4 \ell + 3 \right)}{R^2 \rho^2} + \frac{6}{R^2} -\frac{3 \rho^2}{4 R^2 r^2} \right) \psi^{\ell}_{\mathcal{G},+}(\rho) = 0.
\end{align} 
The second-order differential equations can be written in a more compact and convenient form in the following way, 
\begin{align}
\label{eqnfermionspositive}
\begin{aligned}
&\left[\frac{r^2}{R^2} \partial_{\rho}^2 + \frac{1}{R^2} ~ \left(3 \rho + 3 \frac{r^2}{\rho} \right) \partial_{\rho} + \frac{M^2 R^2}{r^2} + \frac{1}{R^2} \left( 4 + 2 \ell - \frac{r^2}{\rho^2} \left(\ell + \frac{3}{2} \right)  \right) \gamma^{\rho} \right. \\
&\left. + \frac{1}{R^2} \left( -\frac{3 \rho^2}{4 r^2} + 2 - 2 \ell \right) - \frac{r^2}{R^2 \rho^2} \left(\ell^2 + 3 \left( \ell + \frac{1}{2} \right) \right) \right] \psi^\ell_{\mathcal{G}}(\rho) = 0\,,
\end{aligned}
\end{align}
where $k^2=-M^2$. Note this form can also be obtained directly from the first-order formulation of the fluctuations equations, see (\ref{eq: 1st order dirac eq}), by acting upon that equation with $(r \gamma^{\rho} \partial_{\rho} + \frac{1}{r} \gamma^{\nu}  \partial_{\nu})$, i.e. with the first two terms of \eqref{eq: dirac_operator_need_a_label}. In this approach, we need to manipulate (\ref{eq: 1st order dirac eq}) slightly to simplify some terms. As an example, we consider the following expression that is useful for the computations below,  
\begin{align}
\label{eq: how to get rid of gamma_mu}
R^2 \frac{\rho}{r^2} \gamma^{\mu} \partial_{\mu} \hat{\psi}_{\mathcal{F},\mathcal{G}}^\ell = - R^2  \left(\rho \gamma^{\rho} \partial_{\rho} +\frac{1}{2} \left( \frac{\rho^2}{r^2} + 3 \right) \gamma^{\rho} + \frac{\rho^2}{r^2}  \pm \left(\ell + \frac{3}{2} \right)\right) \hat{\psi}_{\mathcal{F},\mathcal{G}}^\ell\,.
\end{align}
Using this and $\gamma^\mu \gamma^\nu \partial_{\mu}\partial_\nu=\eta^{\mu\nu}\partial_\mu\partial_\nu$ we may avoid any explicit appearance of $\gamma^\mu$ in the second order differential equation for $\hat{\psi}^{\ell}_{\mathcal{F,G}}$. For the $\mathcal{G}$ modes, those derived from the positive eigenvalue on the sphere, after inserting the plane wave ansatz \eqref{eq: plane wave spinor}, this again leads to (\ref{eqnfermionspositive}).  

In order to obtain the equivalent expressions for the $\mathcal{F}$-modes we need to consider a sign change in the $B$-factor described in ($\ref{eq: A and B factors}$); namely we have 
\begin{align}
\label{eq: B factors for F}
B &= \frac{\rho}{R r} - \frac{r}{R \rho} \left( \ell + \frac{3}{2}
    \right) \, .
\end{align}
The second-order equations of motion for the $\mathcal{F}$-modes we obtain read 
\begin{align}
\label{eqnfermionsnegative}
\begin{aligned}
&\left[\frac{r^2}{R^2} \partial_{\rho}^2 + \frac{1}{R^2} \left(3 \rho + 3 \frac{r^2}{\rho} \right) \partial_{\rho} + \frac{M^2 R^2}{r^2} + \frac{1}{R^2} \left( -2 - 2 \ell + \frac{r^2}{\rho^2} \left(\ell + \frac{3}{2} \right)  \right) \gamma^{\rho} \right.  \\
&\left. + \frac{1}{R^2}  \left( -\frac{3 \rho^2}{4 r^2} + 8 + 2 \ell \right) - \frac{r^2}{R^2 \rho^2} \left(\ell^2 + 3 \left(\ell + \frac{1}{2} \right) \right) \right] \psi^\ell_{\mathcal{F}}(\rho) = 0\,.
\end{aligned}
\end{align}

\subsection{Large \boldmath{$\rho$} limit and holographic map for the ${\cal G}$ modes}
\label{sec: asymptotes}
Let us consider solving the coupled linear equations ($\ref{eq: coupled system}$) near the conformal boundary, $\rho \rightarrow \infty$. In this limit the mixing terms involving the mesino mass vanish. The equations become 
\begin{align}
\begin{aligned}
\left( \rho ~ \partial_{\rho} +  \left( \ell + \frac{9}{2} \right)    \right) \psi^{\ell}_{\mathcal{G},+}(\rho) &= 0, \\
\left(- \rho ~ \partial_{\rho} +  \left( \ell + \frac{1}{2} \right)    \right) \psi^{\ell}_{\mathcal{G},-}(\rho) &= 0,
\end{aligned}
\end{align}
and the solutions are given by 
\begin{align} \label{eq: asymptotics first oder}
\psi^{\ell}_{\mathcal{G},+}(\rho) &\sim c_{1} ~ \rho^{-\left(\ell+9/2 \right)}, &&& \psi^{\ell}_{\mathcal{G},-}(\rho) &\sim c_{2} ~ \rho^{\ell+1/2},
\end{align}
with $c_{1,2}$ the constants of integration. 

On the other hand, we can take the large-$\rho$ limit at the level of the second-order equations of motion, ($\ref{eqnfermionspositive}$). We can expand in the UV ($\rho \rightarrow \infty$) and obtain 
\begin{align}
\begin{aligned}
\left( \partial_{\rho}^2 + \frac{6}{\rho} \partial_{\rho} + \frac{9/4 -\ell^2 - 4 \ell}{\rho^2} \right) \psi_{\mathcal{G},+}(\rho) &= 0, \\
\left( \partial_{\rho}^2 + \frac{6}{\rho} \partial_{\rho}  - \frac{11/4 - \ell^2 - 6 \ell}{\rho^2} \right) \psi_{\mathcal{G},-}(\rho) &= 0,
\end{aligned}
\end{align}
where the $\pm$ refers to the two different eigenvalues of the
$\gamma^{\rho}$ upon acting on the spinor. The solutions are given by, respectively,
\begin{align}
\begin{aligned} \label{2largersolutions}
\psi_{\mathcal{G},+}(\rho) &\sim c_{3} ~ \rho^{-1/2+\ell} + c_{1} ~ \rho^{-9/2-\ell}, \\
\psi_{\mathcal{G},-}(\rho) &\sim c_{2} ~ \rho^{1/2+\ell} + c_{4} ~ \rho^{-11/2-\ell},
\end{aligned}
\end{align}
with $c_{1,2,3,4}$ being constants of integration. Note we have identified $c_{1,2}$ between the solutions in (\ref{eq: asymptotics first oder}) and (\ref{2largersolutions}). Why though are there extra terms in (\ref{2largersolutions}) relative to (\ref{eq: asymptotics first oder})? The answer is that the two second order equations duplicate the data of the first order equations \cite{Laia:2011wf} - the solutions of one are tied to a particular solution of the other at leading order in $M$ and beyond. To see this we must return to the first order equations to link the solutions. In particular we can substitute (\ref{2largersolutions}) into (\ref{eq: first order sltn}) and (\ref{eq: first order sltn second}). For example if we substitute the $\psi_{\mathcal{G},-}$ solution from (\ref{2largersolutions})  into (\ref{eq: first order sltn}) the $c_2$ term vanishes but the remaining term must reproduce the leading term in $\psi_{\mathcal{G},+}$ in (\ref{eq: asymptotics first oder}) fixing $c_3$ in terms of $c_2$. In this way we can fix the solutions of the second order equations to take the asymptotic form
\begin{align}
\begin{aligned} \label{th1}
\psi_{\mathcal{G},+}(\rho) &\sim - \frac{c_{2}R^2 M }{2(2 + \ell) } ~ \rho^{-1/2 + \ell} + c_{1} ~ \rho^{-9/2 - \ell}, \\
\psi_{\mathcal{G},-}(\rho) &\sim c_{2} ~ \rho^{1/2 + \ell} - \frac{R^2 M c_1}{(6+2\ell)} ~ \rho^{-11/2 - \ell}, 
\end{aligned}
\end{align}
which have the same number of degrees of freedom as the solutions of the linearized equation. Note in practice now we can solve just one of the second order equations and extract $c_1$ and $c_2$ from the asymptotics. 

Let us now consider the holographic dictionary for these modes. We associate the integration constants $c_{1,2}$ with the operator ($\mathcal{O}$) and source ($J$) of a dual field theory operator of dimension $\Delta_{\mathcal{G}} = \ell +  \frac{9}{2}\,$. Note the dimensions of the operator and source add to $d=4$, as expected.

The dual field theory operator naively corresponds to a fermionic
bound state of two fermionic quarks and a gaugino of the ${\cal N}=4$
theory ($\psi_q^\dagger \lambda \psi_q$), dressed with adjoint scalars
at non-zero $\ell$. The exact form of the fermionic operators was
found in \cite{Kirsch:2006he} to be
\begin{equation}
\label{eq: G mode FT}
	\mathcal{G}^\ell	
	\sim~
	\bar{\psi}_i\sigma_{ij}^B\lambda_{C}X^\ell \psi_j
	+
	\bar{q}^m X^B_V\lambda_{C}X^\ell q^m\,,
	\quad
	\text{where}
	\quad B,C=1,2\,.
\end{equation}
Here $\psi_i=(\psi,\tilde{\psi}^\dagger)^T$ is the fundamental  spinor and $\lambda_{\alpha C}$ is the adjoint hypermultiplet. $X^\ell$ is a symmetric and traceless operator insertion of $\ell$ adjoint scalars, $X^{\{i_1}\cdots X^{i_\ell\}}$, where $i=4,5,6,7$. $X_V^B$ is a vector and $\sigma^B=(\sigma^1,\sigma^2)$ a doublet of Pauli matrices.

\subsection{Large \boldmath{$\rho$} limit and holographic map for the ${\cal F}$ modes}

The analysis for the ${\cal F}$ modes follows that for the ${\cal G}$ modes.  We now solve the first order equations ($\ref{eq: coupled system}$) with $B$ in (\ref{eq: B factors for F}) and the second order equation (\ref{eqnfermionsnegative}) at large $\rho$ and identify the integration constants. The asymptotic solution takes the form
\begin{align}
\begin{aligned}
\psi_{\mathcal{F},+}(\rho) &\sim c_{2} ~ \rho^{-3/2 + \ell} + \frac{c_1 M R^2}{2 (\ell+1)} ~ \rho^{-7/2-\ell},  \\
\psi_{\mathcal{F},-}(\rho) & \sim \frac{c_2 M R^2}{2 \ell} ~ \rho^{-5/2 +
  \ell} + c_{1}  ~ \rho^{-5/2-\ell}. \label{c1}
\end{aligned}
\end{align}
Here, the dual field theory mesinos are naively bound states of a
scalar and a gaugino. We associate the integration constants $c_{1,2}$
with the operator ($\mathcal{O}$) and source ($J$) of a dual field
theory operator of dimension $\Delta_{\mathcal{F}} = \ell +
\frac{5}{2}\,$. Note the dimensions of the operator and source add to
$d=4$ as expected. The exact form of
the operator given by $c_1$ in \eqref{c1}   was obtained again in \cite{Kirsch:2006he} and
is given by
\begin{equation}
\label{eq: F mode FT}
	\mathcal{F}^\ell
	\sim
	\bar{q}X^\ell\tilde{\psi}_\alpha^\dagger
	+
	\tilde{\psi}_\alpha X^\ell q.
	\end{equation}

\subsection{Supersymmetric mode solutions \& spectra}
\label{sec: 2nd order eom D7}
In order to determine the mesino mass spectra associated to  the $\mathcal{G}$ and $\mathcal{F}$ modes, we now solve the second-order differential equations for $\hat{\psi}_{\mathcal{F},\mathcal{G}}^\ell$ which we constructed above, ($\ref{eqnfermionspositive}$) and (\ref{eqnfermionsnegative}).

In the supersymmetric theory the source should be set strictly to
zero, whilst $\mathcal{O}$ as a linearized perturbation is a free
parameter corresponding to the normalization. For this case there is a
unique solution to \eqref{eqnfermionspositive} that does not have any  complex infinities \cite{Kirsch:2006he},
\begin{align}
\label{eq: G mode solution D7}
\begin{aligned}
\psi^\ell_{\mathcal{G}}(\rho) = 
&\left( {-L^2} \right)^n \left[\frac{\rho^{\ell+1}}{(\rho^2 + L^2)^{n+\ell+\frac{11}{4}}}  ~ _2F_1 \Big(-n,-(n+\ell+2),\ell+3, -\frac{\rho^2}{L^2}  \Big) \alpha_{+} \right.  \\
&  \left. - {R^2 M (\ell +2) \over 2 (\ell+ n +2)(\ell + n+3)} \frac{\rho^{\ell}}{(\rho^2 + L^2)^{n+\ell+\frac{11}{4}}}  ~ _2F_1 \Big(-n,-(n+\ell+3),\ell+2, -\frac{\rho^2}{L^2}  \Big) \alpha_{-} \right]  \\
\end{aligned} 
\end{align}
here we have fixed the coefficients of each term so that we reproduce precisely the large-$\rho$ behaviour in (\ref{th1}) with the source $c_2$ zero.
This solution corresponds to the mass spectrum 
\begin{align}
\label{eq: MG spectrum D7}
M_{\mathcal{G}} = 2 \frac{L}{R^2} \sqrt{(n+\ell+2)(n+\ell+3)}, && n \geq 0\,, && \ell \geq 0\,.
\end{align}

Next we construct the solutions for the $\mathcal{F}$ modes. These
correspond to the minus sign in ($\ref{eq: chi spherical
  harmonic}$). The solutions are obtained from
($\ref{eqnfermionsnegative}$). Again setting the source to zero and
keeping only those solutions that do not have any complex singularities gives  \cite{Kirsch:2006he}
\begin{align}
\label{eq: F mode solution D7}
\begin{aligned}
\psi^{\ell}_{\mathcal{F}}(\rho) = & (-L^2)^n \left[{R^2 M  \over 2} \frac{\rho^{\ell}}{(\rho^2 + L^2)^{n+\ell+\frac{7}{4}}}  ~ _2F_1 \Big(-n,-(n+\ell+1),\ell+2, -\frac{\rho^2}{L^2}  \Big) \alpha_{+}  \right.~  \\
&\left. +\frac{(n+\ell+1)}{ (\ell+1)} \frac{\rho^{\ell + 1}}{(\rho^2 + L^2)^{n+\ell+\frac{7}{4}}}  ~ _2F_1 \Big(-n,-(n+\ell),\ell+3, -\frac{\rho^2}{L^2}  \Big)  \alpha_{-} \right]\, .
\end{aligned}
\end{align}
The near-boundary expansion of these solutions is given by (\ref{c1}) with the source $c_2$ zero
and the corresponding mass spectrum is
\begin{align}
M_{\mathcal{F}} = 2  \frac{L}{R^2} \sqrt{(n+\ell+1)(n+\ell+2)}, && n \geq 0\,, && \ell \geq 0\,.
\end{align}

\subsection{Open string fluctuation-operator mapping}

We have computed the mass spectra of the spin-$1/2$ modes arising in
the massive canonical D$3$/D$7$ system. As was first shown in
\cite{Kruczenski:2003be},  open string excitations of the probe D$7$-brane fit into massive $\mathcal{N}=2$ supermultiplets. While the counting of the states in the super(conformal)multiplets has been performed in the past, in \cite{Aharony:1998xz} and \cite{Kruczenski:2003be}, it is useful test of our results to check the counting. 

For $L \rightarrow 0$, the fundamental hypermultiplet are massless and
the theory is conformal. The modes are in representations of the
SU$(2)_{R} \times SU(2)_{L} \times U(1)_{R}$ labelled by
$\left(j_{1},j_{2} \right)_{s}$, where $j_{1,2}$ is an index denoting
the spin under the SU$(2)_{R,L}$ respectively, and $s$ is the
eigenvalue associated with the group $U(1)_{R}$. The dimension of
chiral primaries is given by the formula $\Delta = 2 j_{1} + s/2$. Two
scalar fields are  associated with the transverse fluctuations of the D7-brane each of which, after a Kaluza-Klein reduction on the three-sphere, will lead to  tower of real scalars, $\phi^{\ell}$, transforming in the $\left( \frac{\ell}{2}, \frac{\ell}{2} \right)_{2}$, with $\ell \in \mathbb{N}_{0}$. The vector field admits a similar expansion, and from the bulk components on the D$7$-brane we obtain a tower of AdS vectors, $A^{\ell}$, transforming in the $\left(\frac{\ell}{2}, \frac{\ell}{2} \right)_{0}$, with $\ell \in \mathbb{N}_{0}$. Finally, from the components of the vector field on the internal manifold we obtain two different Kaluza-Klein towers of real scalar fields, that we call $A^{\ell}_{\pm}$, transforming in the $\left( \frac{\ell \mp 1}{2}, \frac{\ell \pm 1}{2}\right)_{0}$, with $\ell \in \mathbb{N}$. There are also two types of fermions, which upon reduction on the three sphere will give two towers of states transforming in the $\left(\frac{\ell}{2}, \frac{\ell+1}{2} \right)_{1}$ -the $\mathcal{F}$ fermions- and $\left(\frac{\ell+1}{2}, \frac{\ell}{2} \right)_{1}$ -the $\mathcal{G}$ fermions. 

Introducing a mass gap in the probe-brane setup ($L \neq 0$) breaks the U$(1)_{R}$ acting on the two-dimensional plane that is transverse to both the background and the probe branes and the R-symmetry group is just SU$(2)_{R}$. 

The spectra of the modes are degenerate, namely states with the same
$n+ \ell$ have the same mass. It was observed that such is the
case for the D$3$-brane background in the analysis performed in
\cite{Arean:2006pk, Myers:2006qr}. We proceed to  counting the number of states in a given multiplet. Since the theory has a global $\mathcal{N}=2$ supersymmetry the modes should fill massive supermultiplets, and they have to be in the same representation of the copy of SU$(2)$ that is inert under the supercharges. To arrange this we have to appropriately shift the angular quantum number of the sphere, such that all states fall in the same representation of the SU$(2)_{L}$. This is shown in Table \ref{tablecount}.

\begin{table}[h] \begin{center}
\begin{tabular}{ |c|c|c|c|}
\hline
 Modes & Fluctuation & Representations & Shifted $\ell$  \\
 \hline
  	{$2$ real scalars} & transverse oscillations & $\left(\frac{\ell}{2}, \frac{\ell}{2} \right)$  & $\left(\frac{\ell}{2}, \frac{\ell}{2} \right)$ \\ 
 \hline
 	{$1$ real scalar} & Type $I_{+}$ fluctuations & $\left(\frac{\ell-1}{2}, \frac{\ell+1}{2} \right)$  & $\left(\frac{\ell-2}{2}, \frac{\ell}{2} \right)$ \\ 
 	\hline
 	{$1$ real scalar} & Type $I_{-}$ fluctuations & $\left(\frac{\ell+1}{2}, \frac{\ell-1}{2} \right)$  & $\left(\frac{\ell+2}{2}, \frac{\ell}{2} \right)$ \\ 
 	\hline
 	{$1$ vector} & Type $II$ fluctuations & $\left(\frac{\ell}{2}, \frac{\ell}{2} \right)$  & $\left(\frac{\ell}{2}, \frac{\ell}{2} \right)$ \\ 
 	\hline
 	{$1$ real scalar} & Type $III$ fluctuations & $\left(\frac{\ell}{2}, \frac{\ell}{2} \right)$  & $\left(\frac{\ell}{2}, \frac{\ell}{2} \right)$ \\ 
 	\hline
 	{$1$ Dirac fermion} & Type $\mathcal{F}$ fluctuations & $\left(\frac{\ell+1}{2}, \frac{\ell}{2} \right)$  & $\left(\frac{\ell+1}{2}, \frac{\ell}{2} \right)$ \\ 
 	\hline
 	{$1$ Dirac fermion} & Type $\mathcal{G}$ fluctuations & $\left(\frac{\ell}{2}, \frac{\ell+1}{2} \right)$  & $\left(\frac{\ell-1}{2}, \frac{\ell}{2} \right)$ \\ 
 	\hline
\end{tabular}
\end{center}
\caption{The origin, degrees of freedom and quantum numbers of the fermionic and bosonic states of the ${\cal N}=2$ multiplets of mesinos.}
  \label{tablecount}
\end{table}

Moreover, we have to account for the degeneracy under the SU$(2)_{R}$: we count the degrees of freedom of a given state and then multiply by $\left(2 j_{1} +1 \right)$. Then, the number of bosonic components in a given multiplet for a fixed value of $\ell$ is equal to 
\begin{align}
1 \left( 2 \left( \frac{\ell}{2}+1 \right) + 1 \right) + 6\left(2 \cdot \frac{\ell}{2} +1\right) + 1\left(2 \left(\frac{\ell}{2}-1\right)+1\right)  
\end{align}
and  the number of states for the spin-$1/2$ components in the same multiplet is given by 
\begin{align}
4 \left(2 \frac{\ell+1}{2} +1 \right) + 4 \left(2 \frac{\ell-1}{2} +1 \right) 
\end{align}
For the $\ell=0$ multiplet, we obtain eight bosonic degrees of freedom and an equal number of fermionic states.

\section{Numerically solving for the SUSY spectrum}
Above we have presented closed form solutions to the equations of motion for the fermionic fluctuations \eqref{eqnfermionspositive}, \eqref{eqnfermionsnegative}. Here we present a numerical approach to solving these equations which we will use in  section 5 when we need to find the spectrum in cases where the source for the fermionic operator does not vanish. 

To demonstrate the method, we consider the ${\cal G}$ modes  and we will just concentrate on the $n=0, \ell = 0$ and $n=1, \ell = 0$ cases. We need to solve \eqref{eq: second order psi minus} (or equally we could solve (\ref{eq: second order psi plus}) which contains the same information as we have discussed). We have seen the solution of the differential equations near the boundary, however shooting from the IR to the UV looking for normalizability of the solutions is a much less numerically intensive procedure. 

\begin{figure}[H]
\centering
\includegraphics[scale=0.5]{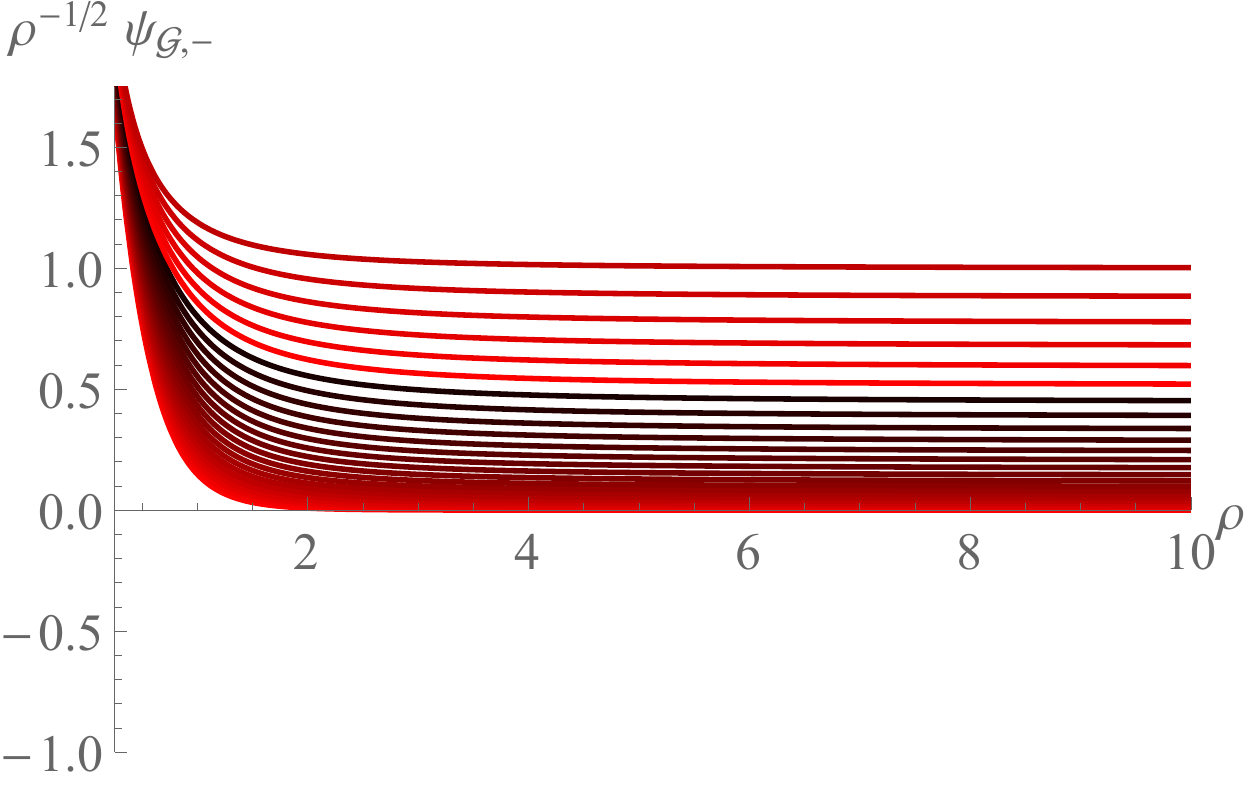} 
\hfill
\includegraphics[scale=0.5]{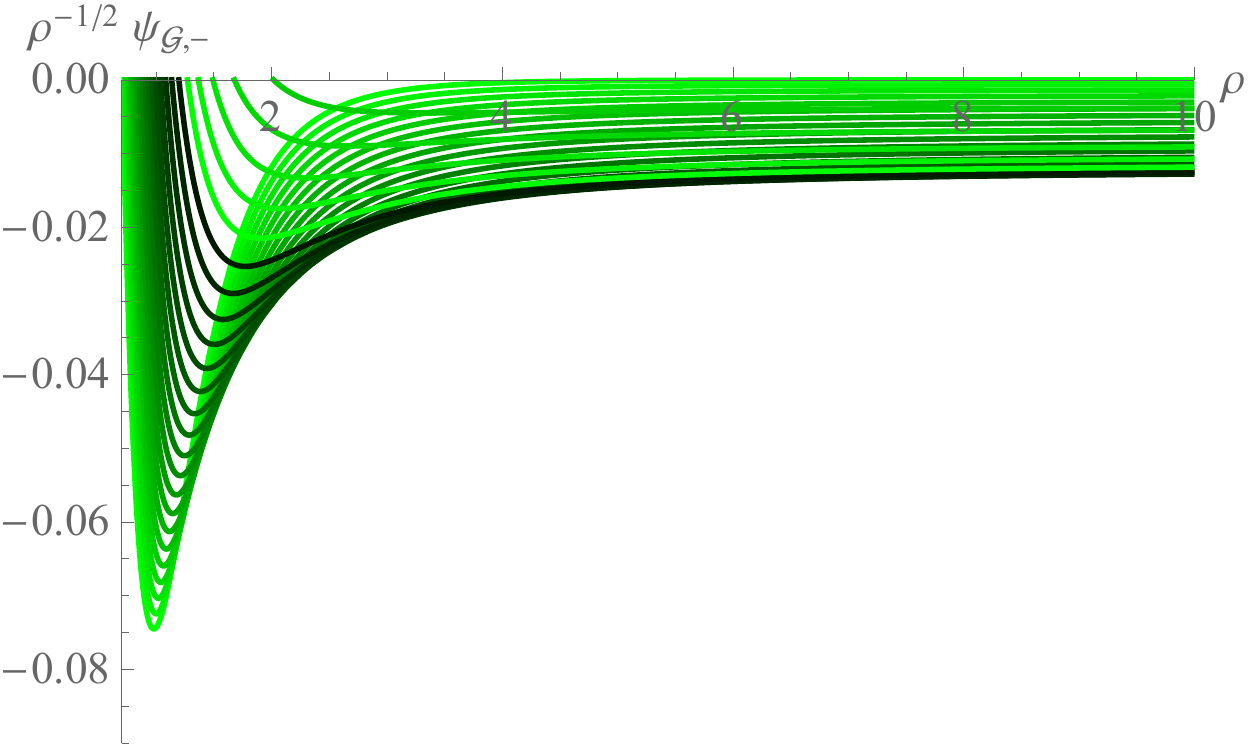} 
\caption{\label{gmesinosflowd7}  Shooting from the IR to the UV for different values of $M^2$ in \eqref{eq: second order psi minus}  for the $\mathcal{G,-}$ type mesinos, using the boundary conditions in \eqref{ircondG}. The left plot shows the results for $\rho^{-1/2} \psi_{\mathcal{G},-}$ for the ground state ($n=\ell=0$) starting from $M^2=0$ and proceeding with steps of one to $M^2=24$ and the right plot corresponds to the first excited state ($n=1, \ell=0$) starting from $M^2=25$ and proceeding with steps of one to $M^2=48$. The solutions relevant to the supersymmetric theory asymptote to zero where the source $J$ vanishes.}
\end{figure}

We expand the analytic solutions to obtain their IR scaling behaviour
and find that
\begin{align} \label{ircondG}
\begin{aligned}
\psi_{\mathcal{G},+}(\rho) &\sim  \rho^{\ell+1}, &&& \partial_{\rho} \psi_{\mathcal{G},+}(\rho) &\sim (\ell+1) \rho^{\ell}, \\
\psi_{\mathcal{G},-}(\rho) & \sim \rho^{\ell}, &&& \partial_{\rho} \psi_{\mathcal{G},-}(\rho) &\sim \ell \rho^{\ell-1}.
\end{aligned}
\end{align}
Thus for $\psi_{\mathcal{G},-}$ we may use the shooting technique
with \eqref{eq: second order psi minus} and for the $\ell=0$ state the boundary conditions
$\psi_{\mathcal{G},-}(0)=1$, $\psi^{'}_{\mathcal{G},-}(0)=0$ to seek
solutions that asymptote to the source $J=0$ in the UV. We recall that
the solution takes the asymptotic form
\begin{align} \label{Guv}
\begin{aligned} 
\psi_{\mathcal{G},+}(\rho) &\sim - {J R^2 M \over 2(2+\ell)}  ~ \rho^{-1/2+\ell} + {\cal O} ~ \rho^{-9/2-\ell}, \\
\psi_{\mathcal{G},-}(\rho) &\sim J ~ \rho^{1/2+\ell} - { {\cal O} R^2 M \over (6 + 2 \ell)} ~ \rho^{-11/2-\ell}, 
\end{aligned}
\end{align}
where $\cal{O}$ is the operator value (we have absorbed factors of $R$ into $M$ for the numerical analysis).  We find it most helpful to
plot $\rho^{-1/2} ~ \psi_{\mathcal{G},-}(\rho)$ since this asymptotes
to $J$. The procedure is simply to shoot out tuning $M^2$ so that
$J=0$ in the UV. In this way, it is straightforward to numerically reproduce the analytic solutions in section 3.7 - we have been able to straightforwardly reproduce the value of $M^2$ of the analytic spectrum numerically to three decimal places. In Figure \ref{gmesinosflowd7} we show this process in action, plotting the solutions for different $M^2$.

\begin{figure}[H]
\centering
\includegraphics[scale=0.5]{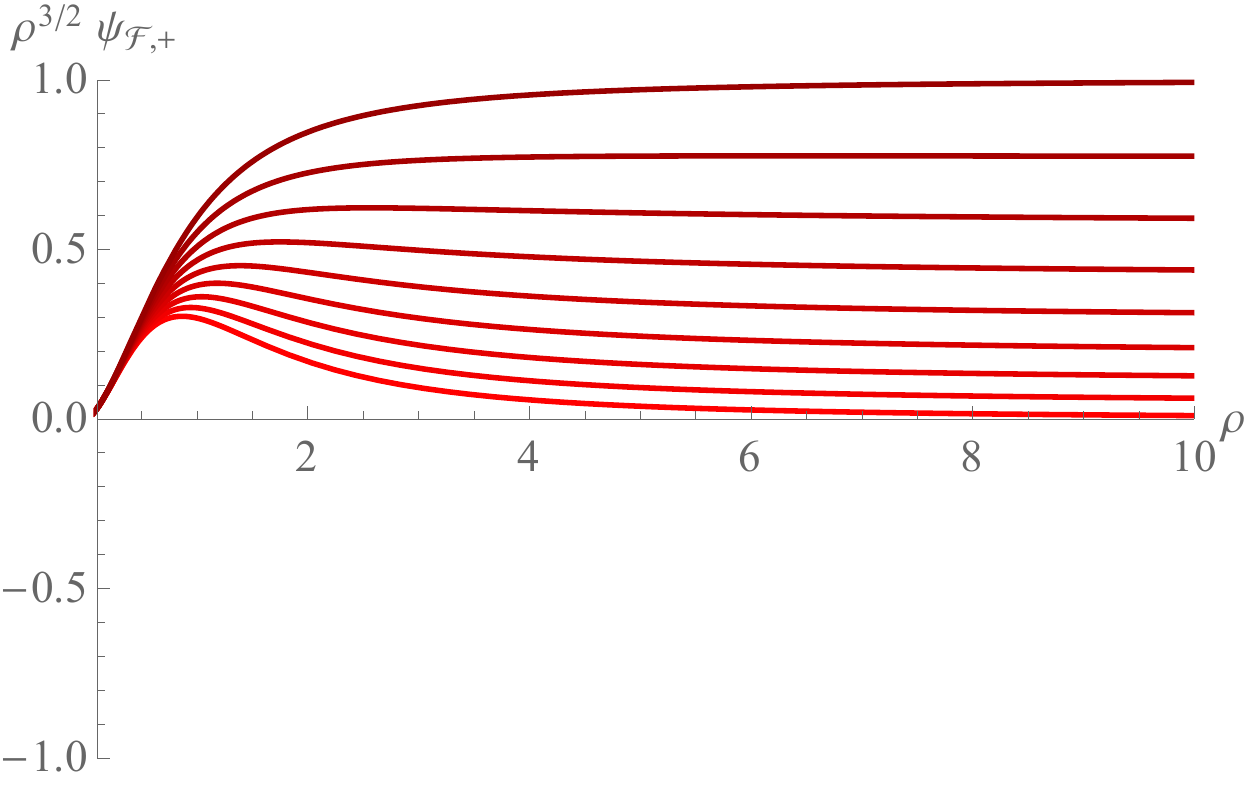} 
\hfill
\includegraphics[scale=0.5]{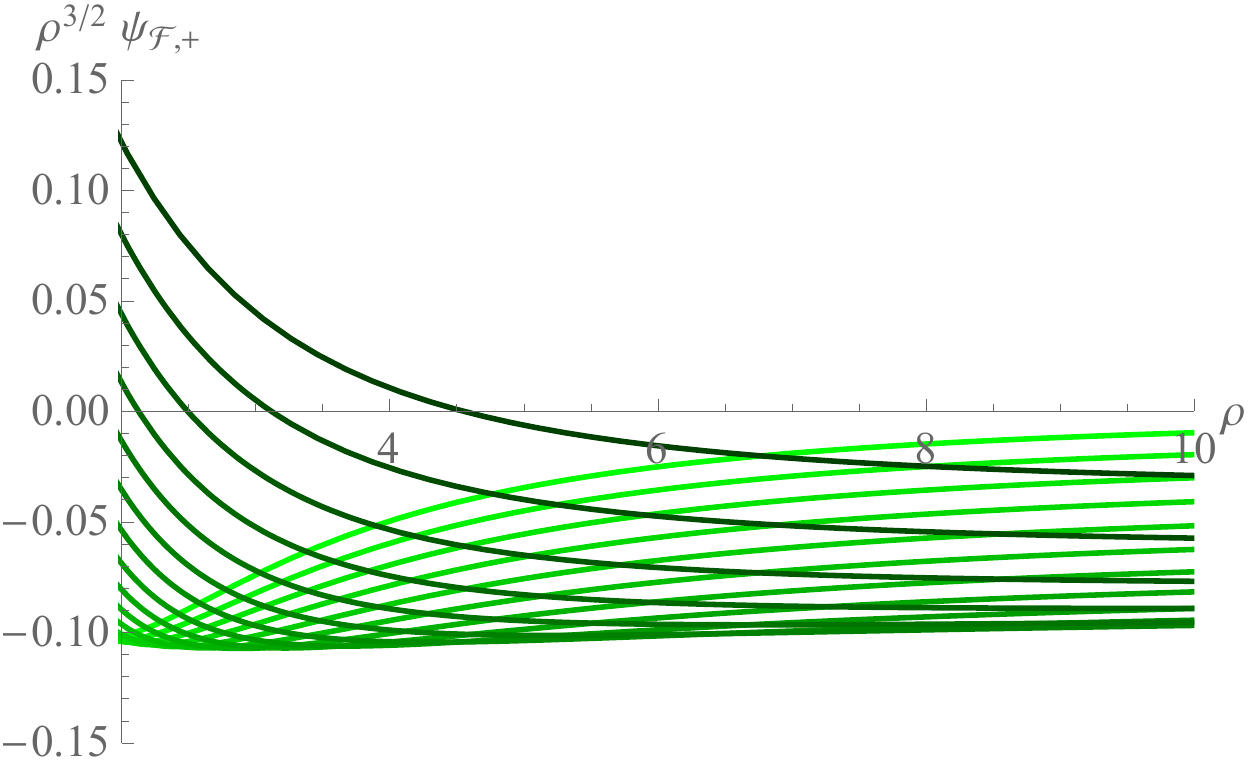} 
\caption{\label{fmesinosflowd7} 
Shooting from the IR to the UV for different values of $M^2$ in (\ref{eqnfermionsnegative})  for the $\mathcal{F,+}$ type mesinos, using the boundary conditions in \eqref{ircondF}. The left plot shows the results for $\rho^{3/2} \psi_{\mathcal{F},+}$ for the ground state ($n=\ell=0$) starting from $M^2=0$ and proceeding with steps of one to $M^2=8$ and the right plot corresponds to the first excited state ($n=1, \ell=0$) starting from $M^2=9$ and proceeding with steps of one to $M^2=24$. The solutions relevant to the supersymmetric theory asymptote to zero where the source $J$ vanishes.}
\end{figure}

We repeat the analysis for the $\mathcal{F}$-modes by solving (\ref{eqnfermionsnegative}). This time we choose to study the differential equation associated with the positive eigenvalue of the chiral $\gamma$-matrix. The IR scaling behaviour here is
\begin{align} \label{ircondF}
\begin{aligned}
\psi_{\mathcal{F},+}(\rho) & \sim \rho^{\ell}, &&& \partial_{\rho} \psi_{\mathcal{F},-}(\rho) &\sim \ell \rho^{\ell-1}, \\
\psi_{\mathcal{F},-}(\rho) &\sim  \rho^{\ell+1}, &&& \partial_{\rho} \psi_{\mathcal{F},+}(\rho) &\sim (\ell+1) \rho^{\ell}, 
\end{aligned}
\end{align}
and the UV asymptotics are
\begin{align}
\begin{aligned}
\psi_{\mathcal{F},+}(\rho) &\sim J ~ \rho^{-3/2 + \ell} + \frac{\mathcal{O}  M R^2}{2(\ell+1)} ~ \rho^{-7/2-\ell},  \\
\psi_{\mathcal{F},-}(\rho) & \sim \frac{J M R^2}{2 \ell} ~ \rho^{-5/2 + \ell} + \mathcal{O}   ~ \rho^{-5/2-\ell}.
\end{aligned}
\end{align}
We solve for $\psi_{\mathcal{F},+}(\rho)$ shooting out from $\psi_{\mathcal{F},+}(0) = 1$, $\psi'_{\mathcal{F},+}(0) = 0$ and seek solutions where $J=0$. It is helpful to plot $\rho^{3/2} ~ \psi_{\mathcal{F},+}(\rho)$ which asymptotes to $J$. Again the supersymmetric states are easily recovered - we show the process in Figure \ref{fmesinosflowd7}.

\section{Double-trace boundary deformations in the D3/D7 system} \label{doubletraced3d7}

So far we have explored the fermionic bound states of the supersymmetric ${\cal N}=2$ gauge theory dual to the D3/ probe D7 system. Our motivation is to find holographic models that give rise to anomalously light fermionic bound states, as required in composite Higgs models. 
What we have seen though is that the spectrum of the supersymmetric
brane models is characterized by the scale $m_q/
\sqrt{\lambda_{YM}}$. As the  't Hooft coupling of the gauge theory grows large, this scale is small relative to the bare quark mass, but it nevertheless sets an intrinsic scale for the strong dynamics. All states lie near that scale, up to order one numerical numbers. This of course has been known for many years, since supersymmetry ties the fermionic bound states to the mesonic bound state masses computed in \cite{Kruczenski:2003be}.

How can we then obtain a baryonic bound state (denoted generically by $\Psi_B$ associated with an operator ${\cal O}_B$), to be light relative to that scale? We wish to explore an answer to that question which consists of including a higher dimension operator in the field theory. These higher dimension operators should be associated with new physics at a UV scale $\Lambda_{UV}$. The precise form of the operator will be chosen so that it corresponds to a shift in the bound state mass at low energies. Generically the approach is this: we add a term to the field-theory Lagrangian of the form 
\begin{equation}
\label{highdimop}
	\Delta {\cal L}_{UV}
	=
	{g^2 \over  \Lambda_{UV}^p}
	\bar{{\cal O}}_B {\cal O}_B,
\end{equation}
where the power, $p$ of the cut off $\Lambda_{UV}$  determined dependent on the UV dimension of the operator. 
As a very simple model, we assume that this operator leads to an RG flow such that in the IR, the baryon $\Psi_B$ receives a mass shift of the form
\begin{equation}
	\Delta {\cal L}_{IR}
	\propto
	{g^2 m_q^{p+1} \over  \Lambda_{UV}^p} \bar{\Psi}_B \Psi_B\,.
\end{equation}
Here we have assumed that the dynamics that binds the fermions occurs
around the quark mass scale where the conformal symmetry is broken -
hence the $m_q$ term which is present to make the operator of
dimension four in the IR. Naively if this term plays a passive role
only, this could be used for a negative shift in the baryon mass that
could be tuned to reduce the baryonic mass scale. In fact we will see
that such operators show a sort of critical behaviour at large $g$ which is more than just this shift.

To include such an operator, we use Witten's multi-trace prescription \cite{Witten:2001ua}. This essentially says that, if the operator \eqref{highdimop} acquires a vev, then a source is generated with the value
\begin{equation}
	J
	=
	{g^2 \over  \Lambda_{UV}^p} \langle {\cal O}_B \rangle\,. \label{condition}
\end{equation}
This relation is imposed on the holographic field corresponding to the operator at the UV cut-off $\rho = \Lambda_{UV}$ - there is thus a large $\rho$  boundary of the dual space.
In practice one just finds solutions with different source-operator combinations and computes $g^2$ at the scale $\Lambda_{UV}$. We have done most of the work for this process in previous sections.

\subsection{An explicit example - the $\ell=0$ ${\cal G}$ mode}

Let us now study an explicit example. We are interested in driving the
mass of one of the mesinos of the ${\cal N}=2$ gauge theory described
by the D3/probe D7 system much lighter than the characteristic scale
$m_q/\sqrt{\lambda_{YM}}$. Let us pick on the lightest $\ell=0,n=0$
${\cal G}$-type mesino discussed above. In particular the masses of
this state are found by solving ($\eqref{eqnfermionspositive}$),  
\begin{align}
\begin{aligned}
&\left[\frac{r^2}{R^2} \partial_{\rho}^2 + \frac{1}{R^2} ~ \left(3 \rho + 3 \frac{r^2}{\rho} \right) \partial_{\rho} + \frac{M^2 R^2}{r^2} + \frac{1}{R^2} \left( 4 + 2 \ell - \frac{r^2}{\rho^2} \left(\ell + \frac{3}{2} \right)  \right) \gamma^{\rho} \right. \\
&\left. + \frac{1}{R^2} \left( -\frac{3 \rho^2}{4 r^2} + 2 - 2 \ell \right) - \frac{r^2}{R^2 \rho^2} \left(\ell^2 + 3 \left( \ell + \frac{1}{2} \right) \right) \right] \psi^0_{\mathcal{G}}(\rho) = 0\,,
\end{aligned}
\end{align}
for the supergravity modes corresponding to the $\mathcal{G}$-type mesinos. We will solve for the negative eigenvalue of $\gamma^\rho$. The UV and IR behaviour of the solutions have been determined in \eqref{ircondG}, \eqref{Guv},
\begin{align}
\begin{aligned} \label{last}
\psi_{\mathcal{G},-}(\rho)_{IR} &\sim  1, &&& \partial_{\rho} \psi_{\mathcal{G},-}(\rho)_{IR} &\sim  0, \\
\psi_{\mathcal{G},-}(\rho)_{UV} &\sim J ~ \rho^{1/2} + \frac{ \mathcal{O} R^2 M}{6} ~ \rho^{-11/2}.
\end{aligned}
\end{align}
In section 4 we gave a full numerical prescription to find these solutions. In Figure \ref{gmesinosflowd7} we display the full set of regular solutions for $\psi_{{\cal G}}$ - each line corresponds to a particular mesino mass $M$ and predicts an associated  value of the source $J$ extracted from the UV asymptotics. In the supersymmetric model we rejected any solutions for which $J\neq 0$ but now we will consider the full set.

Remember that in the dual field theory we are looking at states that are associated with the UV operator in (\ref{eq: G mode FT}) - which includes a three fermion bound state. 
Here consider adding, at the scale $\Lambda_{UV}$, the field-theory Lagrangian term
\begin{equation}
\label{highdimop2}
	\Delta {\cal L}_{UV}
	=
	{g^2 \over  \Lambda_{UV}^5}
	\bar{\mathcal{G}}^{0}\mathcal{G}^0\,.
\end{equation}
The IR mesino $\Psi_M$ receives a mass shift of the form
\begin{equation}
	\Delta {\cal L}_{IR}
	\propto
	{g^2 m_q^6 \over  \Lambda_{UV}^5} \bar{\Psi}_M \Psi_M\,.
\end{equation}
Witten's multi-trace prescription \cite{Witten:2001ua} tells us to require of our regular solutions in Figure \ref{gmesinosflowd7}
\begin{equation}
	J
	=
	{g^2 \over  \Lambda_{UV}^5} \langle \mathcal{G}^0 \rangle\,. \label{condition2}
\end{equation}
We have already numerically computed the solutions to the fluctuation equations for different values of the mass by solving \eqref{eqnfermionspositive} for the mode $\psi^0_{{\cal G}-}$, using the shooting method. We obtain the supersymmetric spectrum from these numerical flows by considering the solutions that asymptote to zero for a vanishing source, $J=0$, and disregarding all other numerical flows. Now, we allow for all the different numerical values of $M^2$ and consider the corresponding numerical solutions we obtained by performing the method described above. For each of those cases we then extract ${\cal O}$ from the UV asymptotics in (\ref{last}). 
Here we determine $J$ and ${\cal O}$ at a value of $\rho$ that corresponds to the UV cut-off $\Lambda_{UV}$ (numerically here we pick $\Lambda_{UV}/L=10$ as an example). 

\begin{figure}[H]
\centering
\includegraphics[scale=0.5]{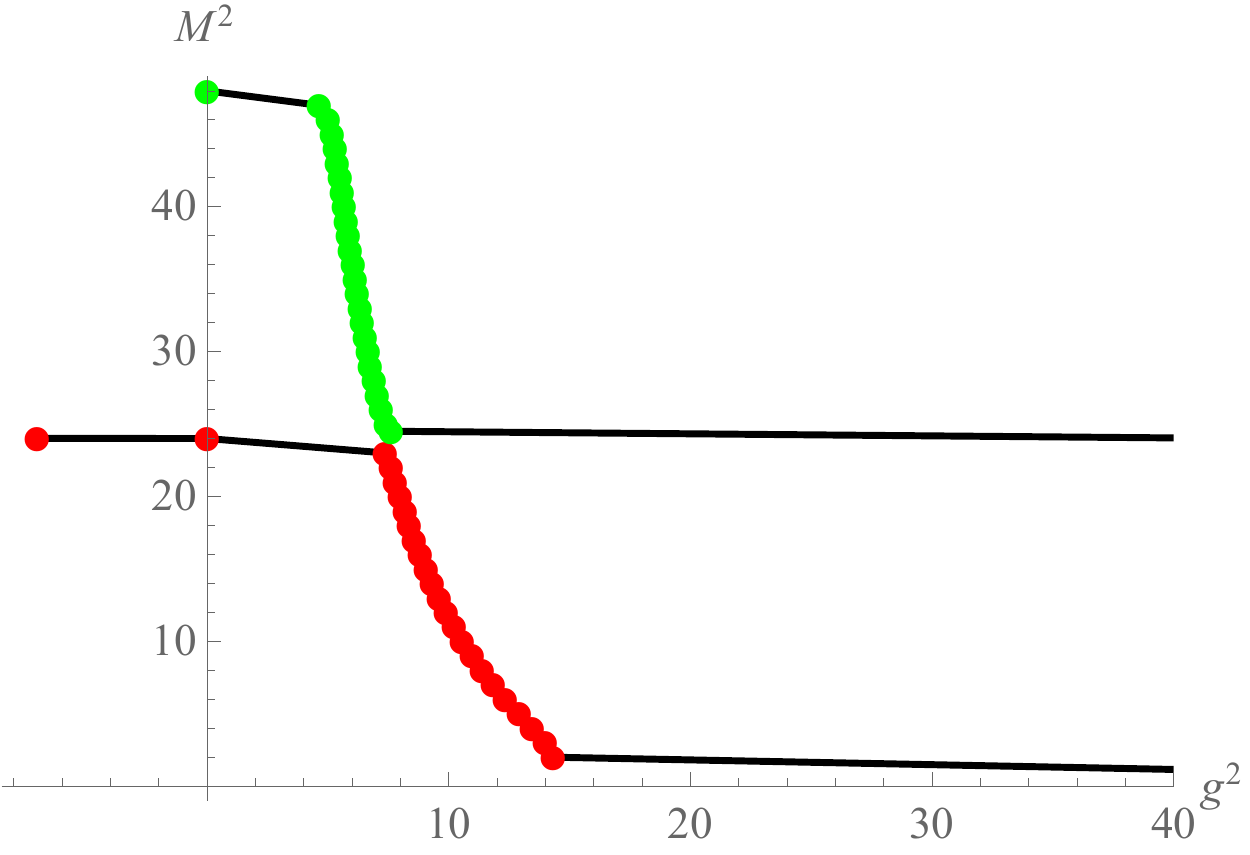} 
\hfill
\includegraphics[scale=0.5]{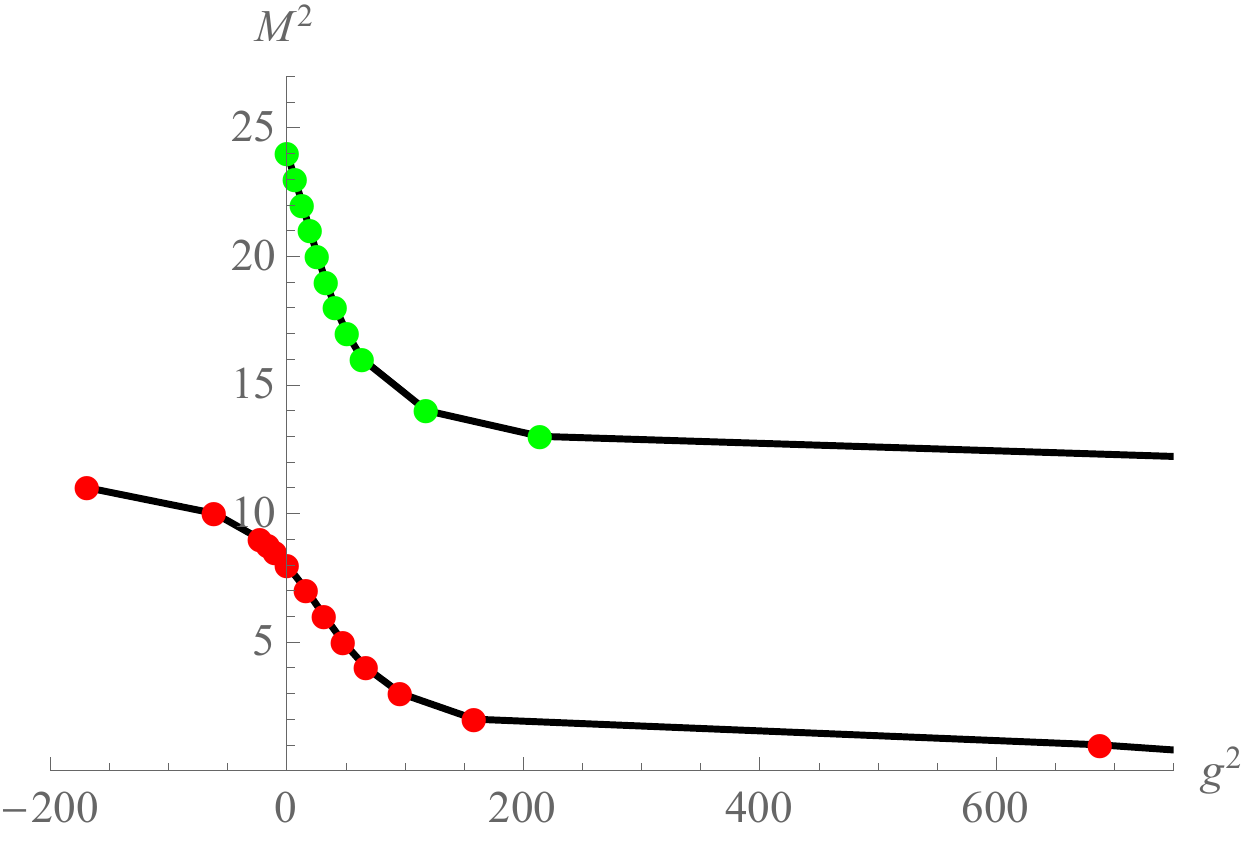} 
\caption{\label{fermionsdoubletrace} D3/D7-brane system: The mesino mass squared $M^2$ as
  function of the coupling strength $g^2$ in units of $L/R^2$(dots are data points whilst the line is to guide the eye) in the presence of the
  double-trace deformation for the $\ell=0$ and $n=0,1$ radially
  excited modes.  The $\mathcal{G}$ fermionic modes are shown on the left
  and the $\mathcal{F}$  modes on the right.  The green points show
  the first, radially excited state getting lighter as the coupling is
  increased, and the red ones show the ground state of the modes.}
\end{figure}

Now we have a series of solutions with $M,J$ and ${\cal O}$ and we may compute the higher dimension operator's coupling $g$ from (\ref{condition2}). 
The result is shown in Fig \ref{fermionsdoubletrace} - it tracks the mass of the mesino against the strength of the coupling $g$.

\noindent The red dots show the lightest state  at each value of $g^2$. As $g^2$ increases from zero, initially the fermionic bound state mass is expected to fall linearly  - the higher dimension operator is  a weak perturbation and the naive analysis applies simply adding a small negative shift to the mesino mass.
In fact it is numerically difficult to extract solutions in this regime because the mesino masses must be very finely tuned close to the supersymmetric value and $g^2$ extracted from the noisy UV asymptotics. The lowest $g^2$ points we extract are consistent with this expectation though. 
Above $g^2=10$ there is a new behaviour though - the mesino mass falls
sharply over a relatively small range of $g^2$. This is suggestive of
the critical behaviour in a Nambu-Jona Lasinio type model where above
a critical value the higher dimension operator is having a major role
in the dynamics. Rather than then driving the mesino mass squared to
zero and negative values though, above $g^2\simeq 15$ the drop in the
mesino mass plateaus before reaching $M^2=0$ only at infinite
coupling. Note that taking the dimensionless $g^2$ large should be an acceptable  theory provided the mesino masses do not rise above the scale $\Lambda_{UV}$ which here they won't because the masses are suppressed by the large 'tHooft coupling.  We believe this region of behaviour is governed by the fact
that fermionic modes cannot condense and so the mass cannot be driven
to become tachyonic. Mathematically, this behaviour follows from the
occurrence of $M$ in the UV solutions for the sub-leading term of the
solution - at $M=0$ if the sub-leading term is non-zero, then the
operator vev is pushed to infinity and hence  also $g^2$ goes to infinity.

Interestingly, adding the term with a negative value of $g^2$ does not
greatly increase the mass of the mesino bound state as one would naively expect - possibly the ${\cal N}=4$ dynamics is already so strong that adding additional strong interactions do not greatly change the dynamics. Such theories have unbounded potentials at the UV cut off in any case. Such a negative $g^2$ can be viewed as a repulsion amongst the fermions; this can also be seen by considering the operator as representing the Feynman diagram of two fermions scattering by the  exchange of a massive gauge boson where repulsion is just a change in the signs.

The behaviour of the green dots that display the first radially
excited state of the $\mathcal{G}^0$ modes is also interesting. These
states too fall in mass as $g^2$ approaches the critical region, but
they saturate at the value of the ground state at $g^2=0$, falling no
lower. The reason is that for each choice of $M^2$, fixing the IR
boundary conditions  fixes the flow - if it flows to a UV boundary
condition corresponding to $g^2=0$,
then that choice of $M^2$ can never occur for any other value of
$g^2$. The expectation therefore is that in this method, only a single
baryonic bound state will be driven to become light, not the full tower of states.  

A similar story can be told for the ${\cal F} $ modes made of a gaugino and a squark - the operator in (\ref{eq: F mode FT}) - and the mass spectrum is also shown in 
Fig \ref{fermionsdoubletrace} as a function of the coupling of the higher dimension operator $g^2/\Lambda \bar{\cal O} {\cal O}$. The same behaviours are observed, namely the lightest state can be driven to have a light mass at intermediate $g^2$ and to zero as $g^2 \rightarrow \infty$. The $n=1, \ell=0$ state falls in mass as the coupling is approaching its critical value, but they saturate at the value of the ground state and never fall lower than that. 

\subsection{Changing the value of the UV cutoff}
\begin{figure}[H]
\centering
\includegraphics[scale=0.5]{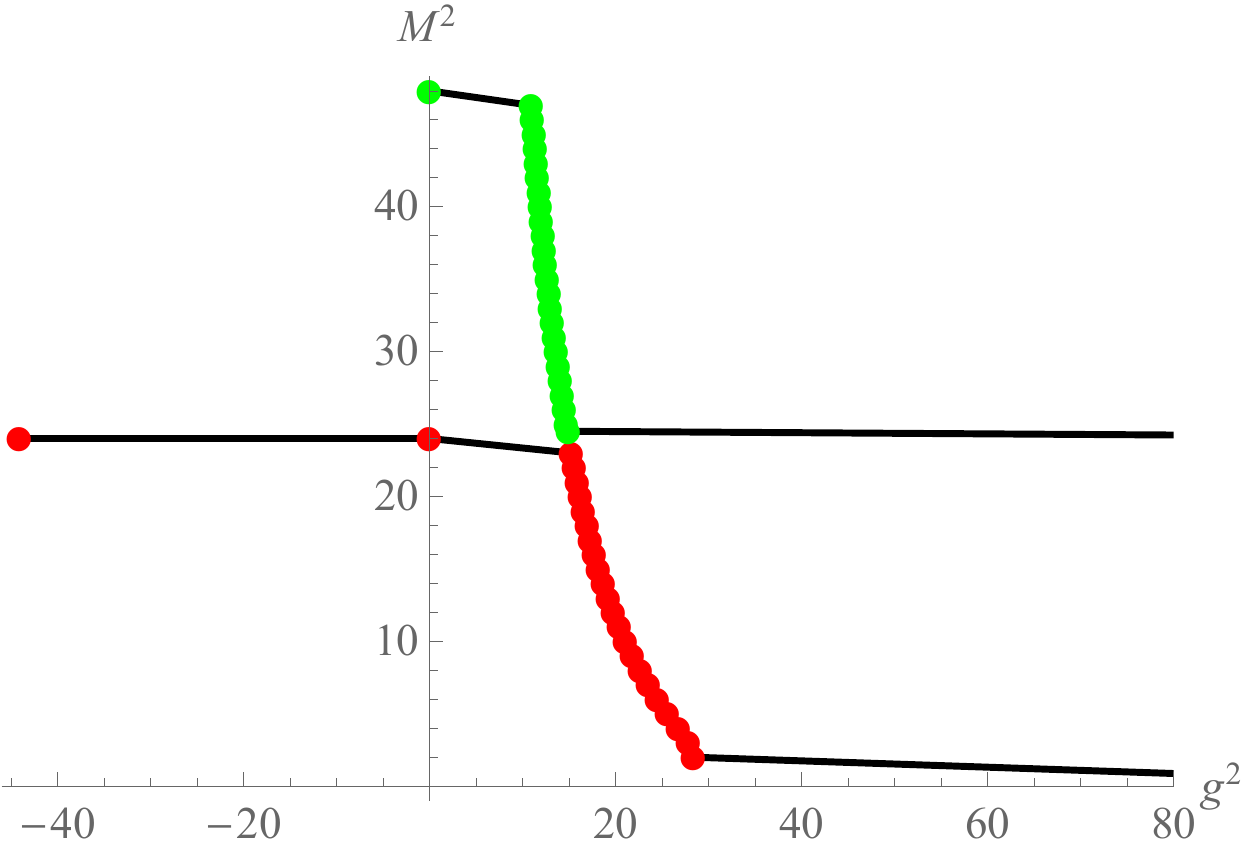} 
\hfill
\includegraphics[scale=0.5]{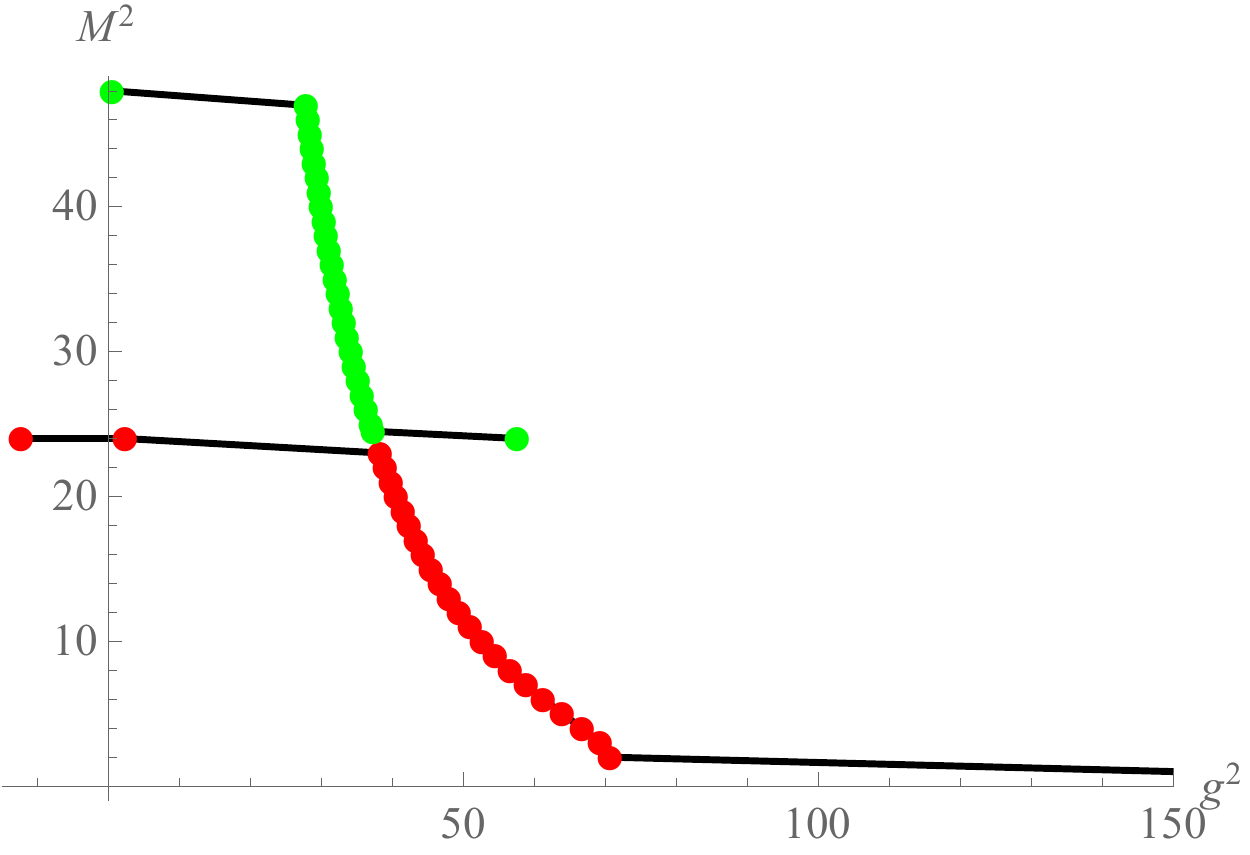} 
\caption{\label{fermionsdoubletracedifferentcutoffs} D3/D7-brane system: The mesino mass squared $M^2$ as
  function of the coupling strength $g^2$ in units of $L/R^2$(dots are data points whilst the line is to guide the eye) in the presence of the
  double-trace deformation for the $\ell=0$ and $n=0,1$ radially
  excited modes.  Here we analyze only the $\mathcal{G}$ fermionic modes. On the left plot we have chosen the value for the UV cutoff to be $20$ and on the right we choose the value $50$. }
\end{figure}

In the previous section  we performed the numerical analysis for the value $\Lambda_{UV}/L=10$. In this section we are interested in the effects that a shift in this cutoff has. We expect the same qualitative behaviour, and indeed this is what we find, see figure \ref{fermionsdoubletracedifferentcutoffs}. The most notable effect is that the  value of $g^2$ where the two branches of the $n=0$ and $n=1$ states nearly meet is raised. We have also estimated the gap in $M^2$ at the point of closest approach and obtain $0.015$, $0.0005$, and $0.00009$ for the three cases $\Lambda_{UV}/L=10,20,50$ respectively suggesting they close together as $\Lambda_{UV}/L$ rises. 

\section{Fermionic fluctuations and higher dimension operators in
  other probe brane systems} \label{sec:intermediate}

The analysis of the fermionic fluctuation in the D$3$/probe D$7$ system above can be extended to a number of other supersymmetric probe brane systems. Here we will restrain ourselves to the D$3$-background of Type IIB, and we will work through these briefly presenting the key equations for the fermionic fluctuations in each case and looking at their response to a higher dimension operator that reduces the mesino masses. The story is very similiar to the D$3$/D$7$ system already discussed, with no alteration in the main considerations. 

We will probe the background generated by a stack of D$3$-branes using D$5$ and D$3$ branes, thus completing all the cases that can be studied analytically. These systems have, as the canonical D$3$/D$7$, eight preserved supercharges. The dual gauge theory in these cases is $4$-dimensional, but the fundamental hypermultiplet has been introduced on a $3$-dimensional and a $2$-dimensional surface respectively for these two cases. We show the agreement with the bosonic sectors of these systems as they were computed in \citep{Arean:2006pk, Myers:2006qr}

\subsection{Fermionic fluctuations in the \boldmath D3/D5 system} \label{d3d5sec}
The difference from the D3/D7 system is that the D$5$-probe wraps an asymptotically AdS$_4 \times$ S$^2$ $\subset$ AdS$_5 \times$ S$^5$. The probe now extends along the $x^0,x^1,x^2$ and $x^4$ directions in the bulk and the Dirac operator on the world-volume of the probe D$5$ acting on the spinor is given by 

\begin{equation}
\label{eq: slashed D in D5}
	\slashed{D}\Psi
	=
	\left(\frac{R}{r} ~ \Gamma^{\mu} ~ \partial_{\mu} + \frac{r}{R} ~  \Gamma^{\rho} ~ \partial_{\rho} + \frac{r}{R \rho} \slashed{\nabla}_{S^{2}} + \frac{1}{2 R} \left(\frac{\rho}{r} +  2 \frac{r}{\rho} \right) \Gamma^{\rho}\right)\Psi
	\,,
\end{equation}	
where, of course, now for the spinor eigenvalues on the sphere we have
to use the analogue of (\ref{eq: chi spherical harmonic}) for a two-dimensional sphere, which reads
$
	\slashed{\nabla}_{S^2}\chi^\pm_\ell
	=
	\pm i\left(\ell
		+
		1
		\right)\chi^\pm_\ell\
$. 

Here we  follow the same procedure that we thoroughly described in the D$3$/D$7$-setup. A minimal way to show how this works in this case is to quote the values of the $A$ and $B$ factors that were introduced in eq.($\ref{eq: A and B factors}$). They read 
\begin{align} \label{eq: new A and B factors}
A &= \frac{1}{2 R} \left( \frac{\rho}{r} + 2 \frac{r}{\rho}  \right),
    \qquad B = \frac{\rho}{R r} \pm \frac{r}{R \rho} \left( \ell + 1
               \right) \, .
\end{align}
We are again led to a system of two first-order coupled differential equations which we showed how to decouple and solve. Let us start by considering the positive sign in eq.($\ref{eq: new A and B factors}$). The corresponding second order differential equation is equal to
\begin{align}
\label{eqnfermionspositived5}
\begin{aligned}
&\left[ \frac{r^2}{R^2} \partial_{\rho}^2 + \frac{1}{R^2} \left(3 \rho + 2 \frac{r^2}{\rho} \right) \partial_{\rho} + \frac{M^2 R^2}{r^2} + \frac{1}{R^2} \left( 3 + 2 \ell - \frac{r^2}{\rho^2} \left(\ell + 1 \right)  \right) \gamma^{\rho} \right. \\
&\left. + \frac{1}{R^2} \left( -\frac{3 \rho^2}{4 r^2} + \frac{3}{2} - 2 \ell \right) - \frac{r^2}{R^2 \rho^2} \left(\ell^2 + 2 \left( \ell + \frac{1}{2} \right) \right) \right] \psi^\ell_{\mathcal{G}}(\rho) = 0\,,
\end{aligned}
\end{align}
We aim at studying the behaviour of the solutions to the above differential equations in the large-$\rho$ expansion. We proceed in a similar way as in the case of the D$3$/probe D$7$ and we obtain 
\begin{align}
\begin{aligned}
\psi_{\mathcal{G},+}(\rho) &\sim \frac{c_{2}M R^2}{2 } ~ \rho^{-1/2 + \ell} + c_{1} ~ \rho^{-7/2 - \ell}, \\
\psi_{\mathcal{G},-}(\rho) &\sim c_{2} ~ \rho^{1/2 + \ell} + \frac{3 c_{1} M R^2}{2} ~ \rho^{-9/2 - \ell}.
\end{aligned}
\end{align}

In order to compute the spectrum of the supersymmetric theory we set the source the source to zero, whilst $O$ as a linearized perturbation is a free parameter corresponding to the normalization. For this case there is a unique solution that has no complex infinities. It is given by
\begin{align}
\label{eq: G mode solution D5}
\begin{aligned}
\psi^\ell_{\mathcal{G}}(\rho) = &\frac{\rho^{\ell+1}}{(\rho^2 + L^2)^{n+\ell+\frac{9}{4}}}  ~ _2F_1 \Big(-n, - \left( n+\ell+\frac{3}{2} \right),\ell+\frac{5}{2}, -\frac{\rho^2}{L^2}  \Big) \alpha_{+} ~  \\
&+d_{\ell n} \frac{\rho^{\ell}}{(\rho^2 + L^2)^{n+\ell+\frac{9}{4}}}  ~ _2F_1 \Big(-n,- \left( n+\ell+\frac{5}{2} \right),\ell+\frac{3}{2}, -\frac{\rho^2}{L^2}  \Big) \alpha_{-},
\end{aligned} 
\end{align}
where, as previously, the spinors $\alpha_{\pm}$ satisfy
\begin{equation}
\label{eq: alpha_pm definition again}
	\gamma^{\rho} \alpha_{\pm} = \pm \alpha_{\pm},
\end{equation}
As we have seen in the D3/probe-D7 analysis in section \ref{sec: 2nd order eom D7} there is a relative $\ell$ and $n$ dependent coefficient between the two hypergeometric solutions, $d_{\ell n}$, which we can evaluate by taking the near-boundary expansion of the exact solution and matching it to the solutions of the asymptotic equations of motion. As we have already given an example for the computation and this coefficient is not relevant for our forthcoming analysis we will not repeat the computation here.

The corresponding mass spectrum is given by
\begin{align}
M_{\mathcal{G}} = 2 \frac{L}{R^2}  \sqrt{ \left( n+\ell+\frac{3}{2} \right) \left( n+\ell+\frac{5}{2} \right)}, && n \geq 0\,, && \ell \geq 0\,.
\end{align}
We  see from the above that the conformal dimension of the dual operator is equal to 
$\Delta_{\mathcal{G}} = \ell +  7/2$. This is again consistent with a
$\psi_q^\dagger \lambda \psi_q$ operator since $\psi_q$ has dimension
1 (it is three-dimensional) and $\lambda$ has dimension 3/2.

In analogy to the $\mathcal{G}$ modes, we may construct the solution for the $\mathcal{F}$ modes, which correspond to the minus sign in eq.($\ref{eq: new A and B factors}$). After decoupling the original set of first-order differential equations, we obtain the following second order one,
\begin{align}
\label{eqnfermionsnegatived5}
\begin{aligned}
&\left[ \frac{r^2}{R^2} \partial_{\rho}^2 + \frac{1}{R^2} \left(3 \rho + 2 \frac{r^2}{\rho} \right) \partial_{\rho} + \frac{M^2 R^2}{r^2} + \frac{1}{R^2}  \left( -1 - 2 \ell - \frac{r^2}{\rho^2} \left(\ell + 1 \right)  \right) \gamma^{\rho} \right.  \\
&\left. + \frac{1}{R^2} \left( -\frac{3 \rho^2}{4 r^2} + \frac{11}{2} + 2 \ell \right)- \frac{r^2}{R^2 \rho^2} \left(\ell^2 + 2 \left(\ell + \frac{1}{2} \right) \right) \right] \psi^\ell_{\mathcal{F}}(\rho) = 0\,.
\end{aligned}
\end{align}
We proceed by examining the large-$\rho$ limit of the above equations and their asymptotic solutions. They are 
\begin{align}
\begin{aligned}
\psi_{\mathcal{F},+}(\rho) &\sim c_{2} ~ \rho^{-3/2 + \ell} + \frac{3
  c_{1} M}{2} ~ \rho^{-5/2-\ell}  \, , \\
\psi_{\mathcal{F},-}(\rho) & \sim \frac{c_{2} M }{2} ~ \rho^{-5/2 +
  \ell} + c_{1}  ~ \rho^{-3/2-\ell} \, .
\end{aligned}
\end{align}
Note the dimensions of the operator and source add to $d=3$ as they should.

We now set the source strictly to zero, in order to obtain the supersymmetric spectrum and the supergravity mode solutions associated with these fermionic fluctuations. The solution to the eq.($\ref{eqnfermionsnegatived5}$) is
\begin{align}
\label{eq: F mode solution D5}
\begin{aligned}
\psi^{\ell}_{\mathcal{F}}(\rho) = &\frac{\rho^{\ell}}{(\rho^2 + L^2)^{n+\ell+\frac{5}{4}}}  ~ _2F_1 \Big(-n,- \left( n+\ell + \frac{1}{2} \right), \ell+\frac{3}{2}, -\frac{\rho^2}{L^2}  \Big) \alpha_{+}  ~  \\
&+ d_{\ell n} \frac{\rho^{\ell + 1}}{(\rho^2 + L^2)^{n+\ell+\frac{5}{4}}}  ~ _2F_1 \Big(-n,- \left( n+\ell + \frac{5}{4}\right),\ell+\frac{5}{2}, -\frac{\rho^2}{L^2}  \Big) \alpha_{-},
\end{aligned}
\end{align}
and the corresponding mass spectrum is given by
\begin{align}
M_{\mathcal{F}} = 2 \frac{L}{R^2}  \sqrt{ \left(n+\ell+\frac{1}{2} \right) \left( n+\ell+\frac{3}{2} \right)}, && n \geq 0, && \ell \geq 0.
\end{align}
The conformal dimension of the associated operator is
$\Delta_{\mathcal{F}} = \ell + 3/2$ again consistent with a scalar quark (dimension 1/2 in 3d) fermionic quark (dimension 1 in 3d) bound state.

\subsubsection{Double-trace boundary deformations in the D3/D5 system}

\begin{figure}[H]
\centering
\includegraphics[scale=0.5]{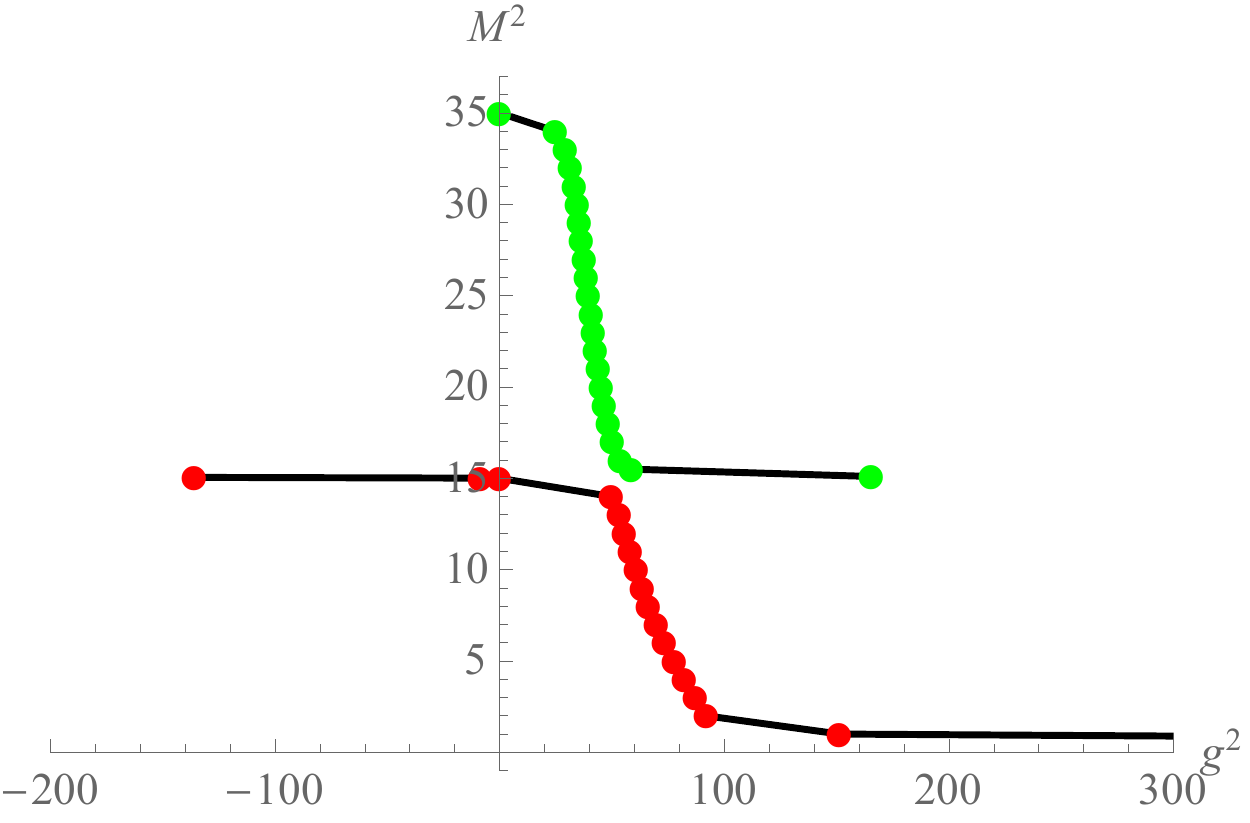} 
\hfill
\includegraphics[scale=0.5]{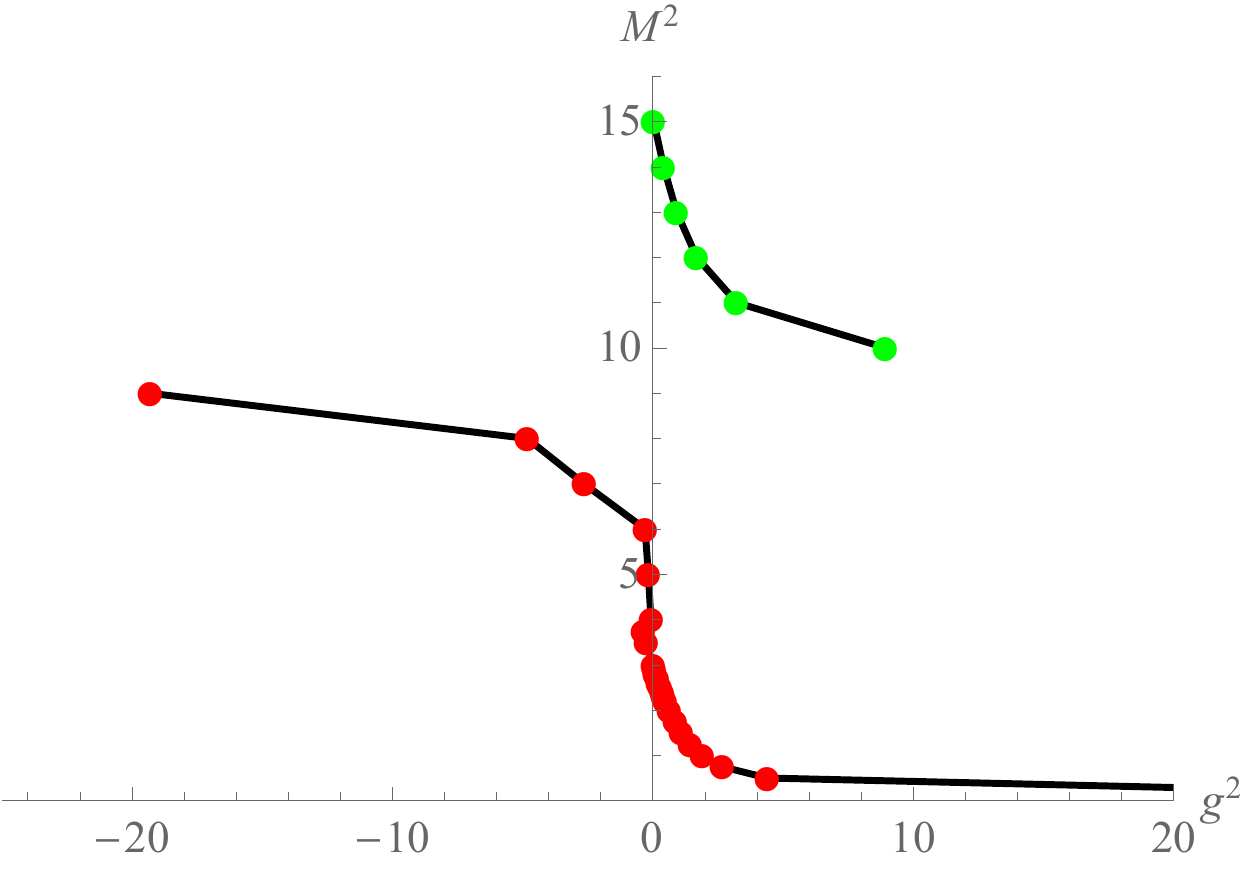} 
\caption{\label{fermionsdoubletraced3d5} D3/D5-brane system: The mesino mass squared $M^2$ in units of $L/R^2$(dots are data points whilst the line is to guide the eye) as
  function of the coupling strength $g^2$  in the presence of the
  double-trace deformation for the $\ell=0$ and $n=0,1$ radially
  excited modes.  The $\mathcal{G}$ fermionic modes are shown on the left
  and the $\mathcal{F}$  modes on the right. The green points show
  the first, radially excited state getting lighter as the coupling is
  increased, and the red ones show the ground state of the modes.}
\end{figure}

Here again we can introduce double trace higher dimension operators that we can use to drive the mesino masses light. The field theory Lagrangian terms are
\begin{equation} \Delta {\cal L}_{\mathcal G}  = {g^2 \over \Lambda_{UV}^4} {\mathcal O}^\dagger_{\mathcal G} {\mathcal O}_{\mathcal G},  \hspace{1cm} 
\Delta {\cal L}_{\mathcal F}  = {g^2} {\mathcal O}^\dagger_{\mathcal F} {\mathcal O}_{\mathcal F}. \end{equation}
Witten's multi-trace prescription tells us then to impose the source  operator relations
\begin{align}
\mathcal{J} = \frac{g^2}{\Lambda^4} \mathcal{O}, \hspace{1cm}
\mathcal{J} = g^2 \mathcal{O}.
\end{align}
As in the D3/D7 numerical studies we shoot from the IR - here to solve (\ref{eqnfermionspositived5}) (\ref{eqnfermionsnegatived5}).  The IR behaviour of the modes are
\begin{align} \label{ircondGd3d5}
\begin{aligned}
\psi_{\mathcal{G},+}(\rho) &\sim  \rho^{\ell+1}, &&& \partial_{\rho} \psi_{\mathcal{G},+}(\rho) &\sim (\ell+1) \rho^{\ell}, \\
\psi_{\mathcal{G},-}(\rho) & \sim \rho^{\ell}, &&& \partial_{\rho} \psi_{\mathcal{G},-}(\rho) &\sim \ell \rho^{\ell-1},
\end{aligned}
\end{align}
and of course similar analysis can be performed for the $\mathcal{F}$-type mesinos. We obtain 
\begin{align} \label{ircondFd3d5}
\begin{aligned}
\psi_{\mathcal{F},+}(\rho) & \sim \rho^{\ell}, &&& \partial_{\rho} \psi_{\mathcal{G},-}(\rho) &\sim \ell \rho^{\ell-1},\\
\psi_{\mathcal{F},-}(\rho) &\sim  \rho^{\ell+1}, &&& \partial_{\rho} \psi_{\mathcal{G},+}(\rho) &\sim (\ell+1) \rho^{\ell}.
\end{aligned}
\end{align}
We now perform the shooting from $\rho=0$ with these conditions for all values of $M^2$ and determine $J$ and ${\mathcal O}$ from the UV asymptotics. 
In figure \ref{fermionsdoubletraced3d5} we show the relation between the coupling and the mass as we make the states lighter. We observe the same features as in the case of the D$3$/D$7$ configuration.

\subsection{Fermionic fluctuations in the \boldmath D3/D3 system} \label{d3d3sec}

Here the D$3$-probe wraps an asymptotically AdS$_3 \times$ S$^1$
$\subset$ AdS$_5 \times$ S$^5$ extending along the $x^0, x^1$
directions, such that it is a one-dimensional defect in the field
theory, as well as the  $x^4$ direction in the bulk. The Dirac operator on the world-volume of the probe D$3$-brane  is equal to
\begin{equation}
\label{eq: slashed D in D3}
	\slashed{D}\Psi
	=
	\left(\frac{R}{r} ~ \Gamma^{\mu} ~ \partial_{\mu} + \frac{r}{R} ~  \Gamma^{\rho} ~ \partial_{\rho} + \frac{r}{R \rho} \slashed{\nabla}_{S^{1}} + \frac{1}{2 R} \left(\frac{\rho}{r} +  \frac{r}{\rho} \right) \Gamma^{\rho}\right)\Psi
	\,,
\end{equation}	
where, of course, now for the spinor eigenvalues on the sphere we have to use the analogue (\ref{eq: chi spherical harmonic}) for a one-dimensional sphere which reads
$
	\slashed{\nabla}_{S^1}\chi^\pm_\ell
	=
	\pm i\left(\ell
		+
		\frac{1}{2}
		\right)\chi^\pm_\ell\
$. 

The relevant values for the $A$ and $B$ factors that were introduced in eq.($\ref{eq: A and B factors}$) in this case read 
\begin{align} \label{eq: again A and B factors}
A &= \frac{1}{2 R} \left( \frac{\rho}{r} + \frac{r}{\rho}  \right), &&& B &= \frac{\rho}{R r} \pm \frac{r}{R \rho} \left( \ell + \frac{1}{2} \right).
\end{align}

We are again led to a system of two first-order coupled differential equations which we showed how to decouple and solve. Let us start by considering the positive sign in eq.($\ref{eq: again A and B factors}$). The corresponding second order differential equation is equal to

\begin{align}
\label{eqnfermionspositived3}
\begin{aligned}
&\left[\frac{r^2}{R^2} \partial_{\rho}^2 + \frac{1}{R^2} \left(3 \rho + \frac{r^2}{\rho} \right) \partial_{\rho} + \frac{M^2 R^2}{r^2} + \frac{1}{R^2} \left( 2 + 2 \ell - \frac{r^2}{\rho^2} \left(\ell + \frac{1}{2} \right)  \right) \gamma^{\rho} \right. \\
&\left. + \frac{1}{R^2}\left( -\frac{3 \rho^2}{4 r^2} + 1 - 2 \ell \right)- \frac{r^2}{R^2 \rho^2} \left(\ell^2 + \ell + \frac{1}{2}  \right) \right] \psi^\ell_{\mathcal{G}}(\rho) = 0\,,
\end{aligned}
\end{align}

We now study the asymptotic behaviour of the solutions to the above differential equations near the boundary. The solutions we obtain are
\begin{align}
\begin{aligned}
\psi_{\mathcal{G},+}(\rho) &\sim \frac{c_{2}M }{2 } ~ \rho^{-1/2 + \ell} + c_{1} ~ \rho^{-5/2 - \ell}, \\
\psi_{\mathcal{G},-}(\rho) &\sim c_{2} ~ \rho^{1/2 + \ell} + \frac{3 c_{1} M}{2} ~ \rho^{-7/2 - \ell}. 
\end{aligned}
\end{align}

In order to compute the mode solutions and the spectrum of the
supersymmetric theory, we set the source to zero. For this case there
is a unique solution that does not have complex infinities. It is given by
\begin{align}
\label{eq: G mode solution D3}
\begin{aligned}
\psi^\ell_{\mathcal{G}}(\rho) = &\frac{\rho^{\ell+1}}{(\rho^2 + L^2)^{n+\ell+\frac{7}{4}}}  ~ _2F_1 \Big(-n, - \left( n+\ell+1 \right),\ell+2, -\frac{\rho^2}{L^2}  \Big) \alpha_{+} ~  \\
&+ d_{\ell n} \frac{\rho^{\ell}}{(\rho^2 + L^2)^{n+\ell+\frac{7}{4}}}  ~ _2F_1 \Big(-n,- \left( n + \ell + 2 \right),\ell + 1, -\frac{\rho^2}{L^2}  \Big) \alpha_{-},
\end{aligned} 
\end{align}
where, as previously, the spinors $\alpha_{\pm}$ satisfy
\begin{equation}
\label{eq: alpha_pm definition again again}
\gamma^{\rho} \alpha_{\pm} = \pm \alpha_{\pm},
\end{equation}
and the $d_{\ell n}$ can be fixed from the UV asymptotics. The corresponding mass spectrum is given by
\begin{align}
M_{\mathcal{G}} = 2 \frac{L}{R^2}  \sqrt{ \left(n + \ell + 1 \right) \left( n + \ell + 2 \right)}, && n \geq 0, && \ell \geq 0.
\end{align}
The conformal dimension of the dual operator being equal to $\Delta_{\mathcal{G}} = \ell +  5/2$ (here $\psi_q$ has dimension 1/2 and the operator is again $\psi_q^\dagger \lambda \psi_q$). 

In analogy to the $\mathcal{G}$ modes, we may construct the solution for the $\mathcal{F}$ modes, which correspond to the minus sign in eq.($\ref{eq: again A and B factors}$). After decoupling the original set of first-order differential equations, we obtain the second order equation
\begin{align}
\label{eqnfermionsnegatived3}
\begin{aligned}
&\left[\frac{r^2}{R^2}  \partial_{\rho}^2 + \frac{1}{R^2} \left(3 \rho + \frac{r^2}{\rho} \right) \partial_{\rho} + \frac{M^2 R^2}{r^2} + \frac{1}{R^2} \left( - 2 \ell + \frac{r^2}{\rho^2} \left(\ell + 1 \right)  \right) \gamma^{\rho} \right.  \\
&\left. + \frac{1}{R^2} \left( -\frac{3 \rho^2}{4 r^2} + 3 + 2 \ell \right) - \frac{r^2}{R^2 \rho^2} \left( \ell^2 +  \ell + \frac{1}{2}  \right) \right] \psi^\ell_{\mathcal{F}}(\rho) = 0\,.
\end{aligned}
\end{align}

We proceed by examining the large-$\rho$ asymptotic expansion of the above equations and their solutions in that limit. They are 
\begin{align}
\begin{aligned}
\psi_{\mathcal{F},+}(\rho) &\sim c_{2} ~ \rho^{-3/2 + \ell} + \frac{3 c_{1} M}{2} ~ \rho^{-5/2-\ell},  \\
\psi_{\mathcal{F},-}(\rho) & \sim \frac{c_{2} M }{2} ~ \rho^{-5/2 + \ell} + c_{1}  ~ \rho^{-1/2-\ell}. 
\end{aligned}
\end{align}
Note the dimensions of the operator and source add to $d=2$ as they should.

We  now  derive the spectrum and the mode solutions of the
supersymmteric theory. In order to do so, we set the source to
zero. For this case there is a unique solution that does not have complex infinities. It is given by
\begin{align}
\label{eq: F mode solution D3}
\begin{aligned}
\psi^{\ell}_{\mathcal{F}}(\rho) = &\frac{\rho^{\ell}}{(\rho^2 + L^2)^{n+\ell+\frac{3}{4}}}  ~ _2F_1 \Big(-n,- \left( n+\ell \right), \ell+1, -\frac{\rho^2}{L^2}  \Big) \alpha_{+}  ~  \\
&+ d_{\ell n} \frac{\rho^{\ell + 1}}{(\rho^2 + L^2)^{n+\ell+\frac{3}{4}}}  ~ _2F_1 \Big(-n,- \left( n+\ell -1 \right), \ell + 2, -\frac{\rho^2}{L^2}  \Big) \alpha_{-},
\end{aligned}
\end{align}
and the corresponding mass spectrum is given by
\begin{align}
M_{\mathcal{F}} = 2 \frac{L}{R^2} \sqrt{ \left(n + \ell \right) \left( n + \ell + 1 \right)} && n \geq 0, && \ell \geq 1.
\end{align}
The conformal dimension of the associated operator is
$\Delta_{\mathcal{F}} = \ell + 1/2$.

\subsubsection{Double-trace boundary deformations in the D3/D3 system}
\begin{figure}[H]
\centering
\includegraphics[scale=0.5]{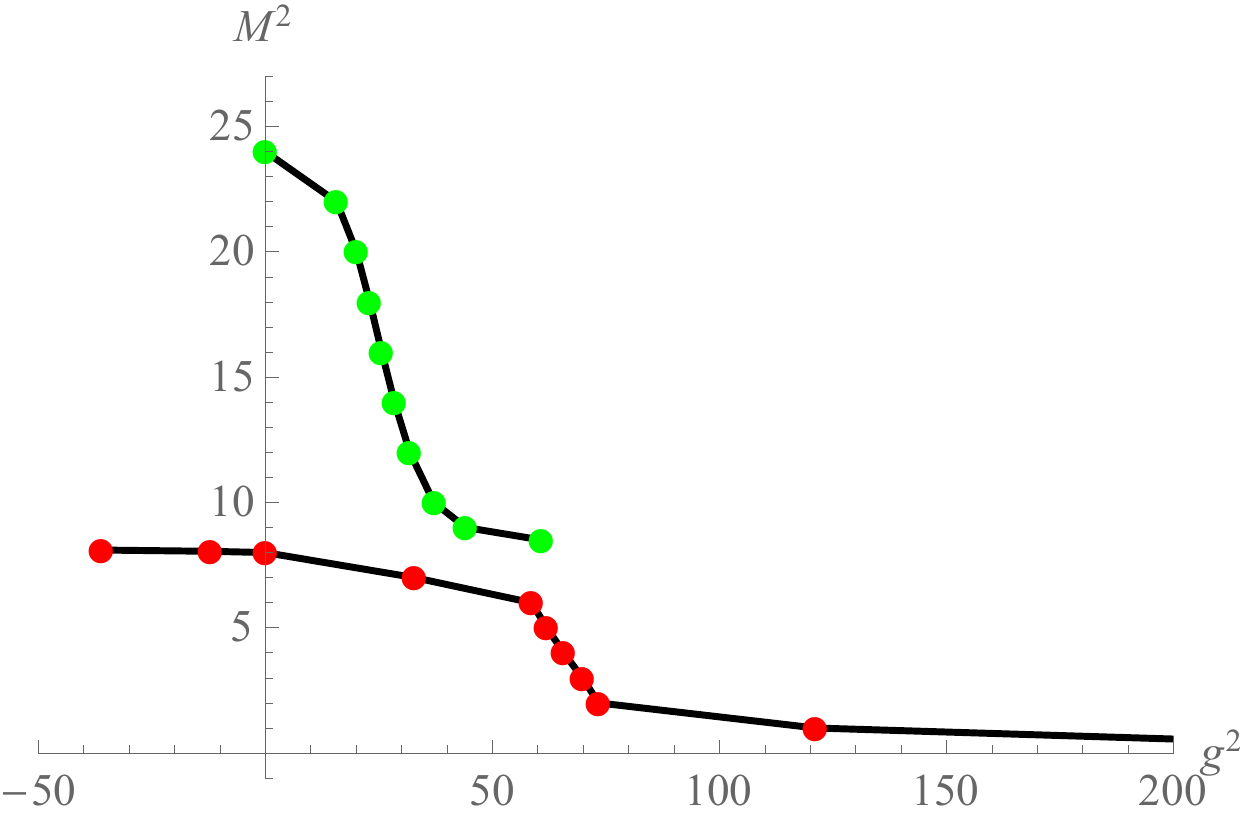} 
\hfill
\includegraphics[scale=0.5]{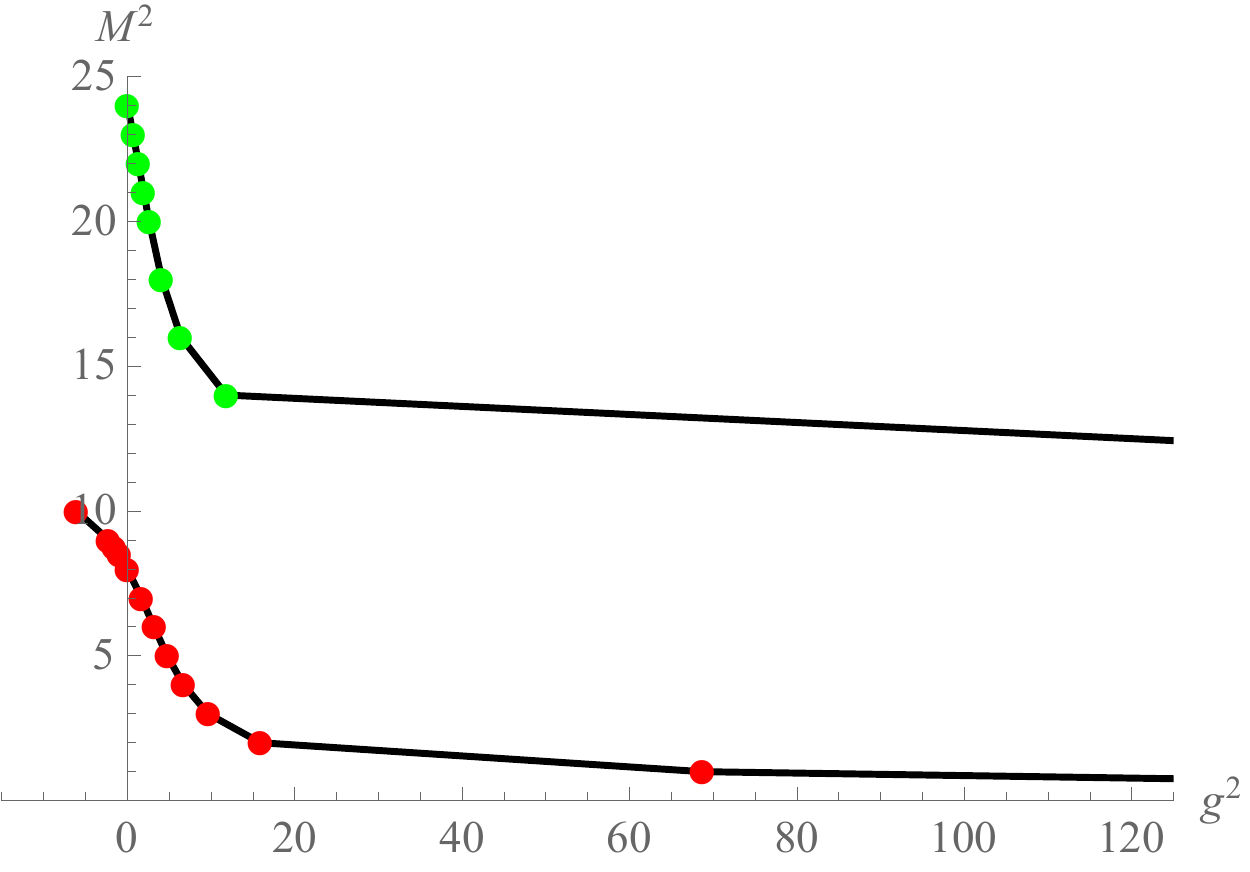} 
\caption{\label{fermionsdoubletraced3d3} 
D3/D3-brane system: The mesino mass squared $M^2$ in units of $L/R^2$(dots are data points whilst the line is to guide the eye) as
  function of the coupling strength $g^2$  in the presence of the
  double-trace deformation for the $\ell=0$ and $n=0,1$ radially
  excited modes for the $\mathcal{G}$ fermionic modes is shown on the left
  and the $\ell=1$ and $n=0,1$ towers of
  states for the $\mathcal{F}$  modes shown on the right.  The green points show
  the first, radially excited state getting lighter as the coupling is
  increased, and the red ones show the ground state of the modes.}
\end{figure}

As in the previous two analyses, we can introduce double-trace higher dimension operators that we can use to drive the mesino masses light. The field theory Lagrangian terms are
\begin{equation} \Delta {\cal L}_{\mathcal G}  = {g^2 \over \Lambda_{UV}^3} {\mathcal O}^\dagger_{\mathcal G} {\mathcal O}_{\mathcal G},  \hspace{1cm} 
\Delta {\cal L}_{\mathcal F}  = {g^2 \over \Lambda_{UV}} {\mathcal O}^\dagger_{\mathcal F} {\mathcal O}_{\mathcal F}. \end{equation}
Witten's multi-trace presecription tells us then to impose the source  operator relations
\begin{align}
\mathcal{J} = \frac{g^2}{\Lambda^3} \mathcal{O}, \hspace{1cm}
\mathcal{J} = \frac{g^2}{\Lambda} \mathcal{O}.
\end{align}
As in the numerical studies in the preceding sections we shoot from
the IR - here to solve (\ref{eqnfermionspositived3}) and (\ref{eqnfermionsnegatived3}).  The IR behaviour of the modes are
\begin{align} \label{ircondGd3d3}
\begin{aligned}
\psi_{\mathcal{G},+}(\rho) &\sim  \rho^{\ell+1}, &&& \partial_{\rho} \psi_{\mathcal{G},+}(\rho) &\sim (\ell+1) \rho^{\ell}, \\
\psi_{\mathcal{G},-}(\rho) & \sim \rho^{\ell}, &&& \partial_{\rho} \psi_{\mathcal{G},-}(\rho) &\sim \ell \rho^{\ell-1},
\end{aligned}
\end{align}
and of course a  similar analysis can be performed for the $\mathcal{F}$-type mesinos. We obtain 
\begin{align} \label{ircondFd3d3}
\begin{aligned}
\psi_{\mathcal{F},+}(\rho) & \sim \rho^{\ell}, &&& \partial_{\rho} \psi_{\mathcal{G},-}(\rho) &\sim \ell \rho^{\ell-1} ,\\
\psi_{\mathcal{F},-}(\rho) &\sim  \rho^{\ell+1}, &&& \partial_{\rho} \psi_{\mathcal{G},+}(\rho) &\sim (\ell+1) \rho^{\ell}.
\end{aligned}
\end{align}
We now  shoot out from $\rho=0$ with these conditions for all values of $M^2$ and determine $J$ and ${\mathcal O}$ from the UV asymptotics. The result of computing the effect of these higher dimension deformations is shown in figure $\ref{fermionsdoubletraced3d3}$.

\section{Conclusions \& summary} \label{sec:conclusions}

We have studied the fermionic fluctuations of massive probe-brane
embeddings in the background generated by a stack of D3-branes. These
are dual to supersymmetric theories that arise from coupling  ${\cal
  N}=4$ $SU(N)$ gauge theory to
hypermultiplets in various dimensions (in four dimensions for the
D7-brane probe
case). We have obtained the supersymmetric mode solutions and the
associated mass spectra. In these cases the probe branes lie flat in
the space and the dimensional reduction of the ten-dimensional spinor
is straightforward. It would be interesting to extend the analysis to
more complex theories. These include the study of fermionic states in
backgrounds that exhibit chiral symmetry breaking,
\cite{Babington:2003vm, Kruczenski:2003uq}. Moreover, one could
consider finite temperature effects due to a black hole in the bulk. A
particularly challenging task would be probes in the presence of 
non-vanishing Kalb-Ramond fields \cite{Filev:2007gb}, where the
description is essentially string theory in a non-commutative
background \cite{Seiberg:1999vs}. In order to perform this analysis
beyond the probe-approximation \cite{Nunez:2010sf}, we need to address the question of the appropriate form for fermionic D-brane action for multiple branes.

We also considered the addition of higher dimension operators of
``baryon squared'' form and the effect they have on the masses of
fermionic bound states. The results are well summarized by Figure
$\ref{fermionsdoubletrace}$ in the D3/D7 system. We have found that
the addition of these higher dimension operators can drive the mesino
masses to light values as compared to the rest of the spectrum. The effect of these
operators is small for small values of the coupling, and the shift in
the mesino mass is linear. For higher values we observed a critical
behaviour with a rapid decrease in the mesino mass. Nevertheless, we
have found that the mode can only be made massless for an infinite
value of the coupling. We have repeated these computations for the
D3/D5 and D3/D3 supersymmetric defect theories to check that this
behaviour is generic to such models. 

Note that the mesino spectra shown in Figures
$\ref{fermionsdoubletrace}$, 4 and 5  suggest an avoided level crossing,
i.e.~asymptotically for very large coupling of the double trace
operator, the mesino mass corresponding to the $n=1$ level approaches
the mesino mass value of the $n=0$ level at vanishing
or repulsive $g^2$. Such a level crossing is known to occur in the
D3/probe D7-brane system if an instanton configuration is considered
in the four D7-brane dimensions perpendicular to the D3-branes
\cite{Erdmenger:2005bj}.  There, for infinite instanton radius, the
meson mass is shifted by two levels as compared to zero instanton
size. This shift was shown to be equivalent to a large gauge
transformation. Here however, the level crossing mechanism is different since
it is triggered by tuning the coupling of a double-trace operator. On
the gravity side, this means the shift occurs in the asymptotic
boundary behaviour of the solutions rather than in the fluctuation
equation of motion itself. 
The
separation between the ground state and first excited state branches
of the curve is presumably controlled by the only dimensionless
parameter, the ratio of the IR mass scale and the UV cut-off scale, $L
/ \Lambda$. It may be instructive  to understand this mechanism in
more detail in the future, for instance by analyzing the underlying
Schr\"odinger equation for the fluctuations.

To conclude, we emphasize again that the higher dimension operators
can be used, by tuning the coupling, to  generate light baryonic
states. Of course in a true model of the UV cut-off physics, it is
unlikely that such an operator would exist in isolation, but our study
shows that in principle such operators could play this role. We are
motivated by Beyond the Standard Model theories, where composite
fermionic or light top partner states are desired. We intend to move
the mechanism displayed here to holographic descriptions of more
phenomenologically appropriate gauge theories in the near future.

\appendix
\section{Notation}

In this paper we use the following index conventions.
Capital latin letters from the middle of the alphabet starting from $M,N,\dots$,
denote ten-dimensional spacetime indices, while capital latin letters starting from $I,J,\dots$ refer to the ten-dimensional Lorentz frame.
Capital letters from the beginning of the alphabet, i.e. $A,B\dots$ are probe brane indices.

Greek lower case letters from the middle of the alphabet, i.e. $\mu,\nu,\dots$,
refer to Minkowski indices, i.e. to the directions of the branes generating
the ten-dimensional curved space time.
Greek lower case letters from the beginning of the alphabet, i.e. $\alpha,\beta,\dots$, denote radial and Minkowski coordinates, i.e. $x^\mu$ and $\rho$. Lower case latin indices $i,j,k,\dots$ are valued on the sphere and the tilded letters $\tilde{m},\tilde{n},\dots$ are the directions transverse to both the background and the probe branes. To simplify notation we do not use separate symbols for curved and flat spacetime indices as it should be clear from the context. In cases that we think that it is not, we provide additional explanations and comments.

We have used the standard conventions of forms, namely a $p$-form is written as 

\begin{align}
A_{p} = \frac{1}{p!} ~ A_{a_{1} \cdots a_{p}} ~ dx^{a_1} \wedge \cdots \wedge dx^{a_p}
\end{align}

Moreover, we consider branes with a positive Chern-Simons term. The above are chosen such that we follow closely the conventions of \cite{Martucci:2005rb} thoughout. \\

\noindent {\bf Acknowledgements:} the authors are grateful for discussions with Werner Porod, James Drummond and Alexander Broll and to Alfonso Ramallo for providing us with his notes on the mesino computation in the supersymmetric theories. NE's
work  was  supported  by  the  STFC  consolidated  grant ST/P000711/1.

\bibliographystyle{JHEP}
\bibliography{mesino_lib}

\providecommand{\href}[2]{#2}\begingroup\raggedright\begin{thebibliography}{10}

\bibitem{Dimopoulos:1980hn}
S.~Dimopoulos, S.~Raby and L.~Susskind, \emph{{Light Composite Fermions}},
  \href{https://doi.org/10.1016/0550-3213(80)90215-1}{\emph{Nucl. Phys.}
  {\bfseries B173} (1980) 208--228}.

\bibitem{Kaplan:1991dc}
D.~B. Kaplan, \emph{{Flavor at SSC energies: A New mechanism for dynamically
  generated fermion masses}},
  \href{https://doi.org/10.1016/S0550-3213(05)80021-5}{\emph{Nucl. Phys.}
  {\bfseries B365} (1991) 259--278}.

\bibitem{Ferretti:2013kya}
G.~Ferretti and D.~Karateev, \emph{{Fermionic UV completions of Composite Higgs
  models}}, \href{https://doi.org/10.1007/JHEP03(2014)077}{\emph{JHEP}
  {\bfseries 03} (2014) 077},
  [\href{https://arxiv.org/abs/1312.5330}{{\ttfamily 1312.5330}}].

\bibitem{Maldacena:1997re}
J.~M. Maldacena, \emph{{The Large N limit of superconformal field theories and
  supergravity}}, \href{https://doi.org/10.1023/A:1026654312961,
  10.4310/ATMP.1998.v2.n2.a1}{\emph{Int. J. Theor. Phys.} {\bfseries 38} (1999)
  1113--1133}, [\href{https://arxiv.org/abs/hep-th/9711200}{{\ttfamily
  hep-th/9711200}}].

\bibitem{Witten:1998qj}
E.~Witten, \emph{{Anti-de Sitter space and holography}},
  \href{https://doi.org/10.4310/ATMP.1998.v2.n2.a2}{\emph{Adv. Theor. Math.
  Phys.} {\bfseries 2} (1998) 253--291},
  [\href{https://arxiv.org/abs/hep-th/9802150}{{\ttfamily hep-th/9802150}}].

\bibitem{Gubser:1998bc}
S.~S. Gubser, I.~R. Klebanov and A.~M. Polyakov, \emph{{Gauge theory
  correlators from noncritical string theory}},
  \href{https://doi.org/10.1016/S0370-2693(98)00377-3}{\emph{Phys. Lett.}
  {\bfseries B428} (1998) 105--114},
  [\href{https://arxiv.org/abs/hep-th/9802109}{{\ttfamily hep-th/9802109}}].

\bibitem{Witten:1998xy}
E.~Witten, \emph{{Baryons and branes in anti-de Sitter space}},
  \href{https://doi.org/10.1088/1126-6708/1998/07/006}{\emph{JHEP} {\bfseries
  07} (1998) 006}, [\href{https://arxiv.org/abs/hep-th/9805112}{{\ttfamily
  hep-th/9805112}}].

\bibitem{Babington:2003vm}
J.~Babington, J.~Erdmenger, N.~J. Evans, Z.~Guralnik and I.~Kirsch,
  \emph{{Chiral symmetry breaking and pions in nonsupersymmetric gauge /
  gravity duals}},
  \href{https://doi.org/10.1103/PhysRevD.69.066007}{\emph{Phys. Rev.}
  {\bfseries D69} (2004) 066007},
  [\href{https://arxiv.org/abs/hep-th/0306018}{{\ttfamily hep-th/0306018}}].

\bibitem{Karch:2002sh}
A.~Karch and E.~Katz, \emph{{Adding flavor to AdS / CFT}},
  \href{https://doi.org/10.1088/1126-6708/2002/06/043}{\emph{JHEP} {\bfseries
  06} (2002) 043}, [\href{https://arxiv.org/abs/hep-th/0205236}{{\ttfamily
  hep-th/0205236}}].

\bibitem{Kruczenski:2003be}
M.~Kruczenski, D.~Mateos, R.~C. Myers and D.~J. Winters, \emph{{Meson
  spectroscopy in AdS / CFT with flavor}},
  \href{https://doi.org/10.1088/1126-6708/2003/07/049}{\emph{JHEP} {\bfseries
  07} (2003) 049}, [\href{https://arxiv.org/abs/hep-th/0304032}{{\ttfamily
  hep-th/0304032}}].

\bibitem{Erdmenger:2007cm}
J.~Erdmenger, N.~Evans, I.~Kirsch and E.~Threlfall, \emph{{Mesons in
  Gauge/Gravity Duals - A Review}},
  \href{https://doi.org/10.1140/epja/i2007-10540-1}{\emph{Eur. Phys. J.}
  {\bfseries A35} (2008) 81--133},
  [\href{https://arxiv.org/abs/0711.4467}{{\ttfamily 0711.4467}}].

\bibitem{Kirsch:2006he}
I.~Kirsch, \emph{{Spectroscopy of fermionic operators in AdS/CFT}},
  \href{https://doi.org/10.1088/1126-6708/2006/09/052}{\emph{JHEP} {\bfseries
  09} (2006) 052}, [\href{https://arxiv.org/abs/hep-th/0607205}{{\ttfamily
  hep-th/0607205}}].

\bibitem{informal}
D.~Are\'an, I.~Kirsch and A.~Ramallo, ``Private notes.'' \ndd.

\bibitem{Faraggi:2011bb}
A.~Faraggi and L.~A. Pando~Zayas, \emph{{The Spectrum of Excitations of
  Holographic Wilson Loops}},
  \href{https://doi.org/10.1007/JHEP05(2011)018}{\emph{JHEP} {\bfseries 05}
  (2011) 018}, [\href{https://arxiv.org/abs/1101.5145}{{\ttfamily 1101.5145}}].

\bibitem{Martucci:2005rb}
L.~Martucci, J.~Rosseel, D.~Van~den Bleeken and A.~Van~Proeyen, \emph{{Dirac
  actions for D-branes on backgrounds with fluxes}},
  \href{https://doi.org/10.1088/0264-9381/22/13/014}{\emph{Class. Quant. Grav.}
  {\bfseries 22} (2005) 2745--2764},
  [\href{https://arxiv.org/abs/hep-th/0504041}{{\ttfamily hep-th/0504041}}].

\bibitem{Laia:2011wf}
J.~N. Laia and D.~Tong, \emph{{Flowing Between Fermionic Fixed Points}},
  \href{https://doi.org/10.1007/JHEP11(2011)131}{\emph{JHEP} {\bfseries 11}
  (2011) 131}, [\href{https://arxiv.org/abs/1108.2216}{{\ttfamily 1108.2216}}].

\bibitem{Witten:2001ua}
E.~Witten, \emph{{Multitrace operators, boundary conditions, and AdS / CFT
  correspondence}},  \href{https://arxiv.org/abs/hep-th/0112258}{{\ttfamily
  hep-th/0112258}}.

\bibitem{Evans:2016yas}
N.~Evans and K.-Y. Kim, \emph{{Holographic Nambu - Jona - Lasinio
  interactions}}, \href{https://doi.org/10.1103/PhysRevD.93.066002}{\emph{Phys.
  Rev.} {\bfseries D93} (2016) 066002},
  [\href{https://arxiv.org/abs/1601.02824}{{\ttfamily 1601.02824}}].

\bibitem{Camporesi:1995fb}
R.~Camporesi and A.~Higuchi, \emph{{On the Eigen functions of the Dirac
  operator on spheres and real hyperbolic spaces}},
  \href{https://doi.org/10.1016/0393-0440(95)00042-9}{\emph{J. Geom. Phys.}
  {\bfseries 20} (1996) 1--18},
  [\href{https://arxiv.org/abs/gr-qc/9505009}{{\ttfamily gr-qc/9505009}}].

\bibitem{Ammon:2010pg}
M.~Ammon, J.~Erdmenger, M.~Kaminski and A.~O'Bannon, \emph{{Fermionic Operator
  Mixing in Holographic p-wave Superfluids}},
  \href{https://doi.org/10.1007/JHEP05(2010)053}{\emph{JHEP} {\bfseries 05}
  (2010) 053}, [\href{https://arxiv.org/abs/1003.1134}{{\ttfamily 1003.1134}}].

\bibitem{Kim:1985ez}
H.~J. Kim, L.~J. Romans and P.~van Nieuwenhuizen, \emph{{The Mass Spectrum of
  Chiral N=2 D=10 Supergravity on S**5}},
  \href{https://doi.org/10.1103/PhysRevD.32.389}{\emph{Phys. Rev.} {\bfseries
  D32} (1985) 389}.

\bibitem{Mueck:1998iz}
W.~Mueck and K.~S. Viswanathan, \emph{{Conformal field theory correlators from
  classical field theory on anti-de Sitter space. 2. Vector and spinor
  fields}}, \href{https://doi.org/10.1103/PhysRevD.58.106006}{\emph{Phys. Rev.}
  {\bfseries D58} (1998) 106006},
  [\href{https://arxiv.org/abs/hep-th/9805145}{{\ttfamily hep-th/9805145}}].

\bibitem{Aharony:1998xz}
O.~Aharony, A.~Fayyazuddin and J.~M. Maldacena, \emph{{The Large N limit of
  N=2, N=1 field theories from three-branes in F theory}},
  \href{https://doi.org/10.1088/1126-6708/1998/07/013}{\emph{JHEP} {\bfseries
  07} (1998) 013}, [\href{https://arxiv.org/abs/hep-th/9806159}{{\ttfamily
  hep-th/9806159}}].

\bibitem{Arean:2006pk}
D.~Are\'an and A.~V. Ramallo, \emph{{Open string modes at brane
  intersections}},
  \href{https://doi.org/10.1088/1126-6708/2006/04/037}{\emph{JHEP} {\bfseries
  04} (2006) 037}, [\href{https://arxiv.org/abs/hep-th/0602174}{{\ttfamily
  hep-th/0602174}}].

\bibitem{Myers:2006qr}
R.~C. Myers and R.~M. Thomson, \emph{{Holographic mesons in various
  dimensions}},
  \href{https://doi.org/10.1088/1126-6708/2006/09/066}{\emph{JHEP} {\bfseries
  09} (2006) 066}, [\href{https://arxiv.org/abs/hep-th/0605017}{{\ttfamily
  hep-th/0605017}}].

\bibitem{Kruczenski:2003uq}
M.~Kruczenski, D.~Mateos, R.~C. Myers and D.~J. Winters, \emph{{Towards a
  holographic dual of large N(c) QCD}},
  \href{https://doi.org/10.1088/1126-6708/2004/05/041}{\emph{JHEP} {\bfseries
  05} (2004) 041}, [\href{https://arxiv.org/abs/hep-th/0311270}{{\ttfamily
  hep-th/0311270}}].

\bibitem{Filev:2007gb}
V.~G. Filev, C.~V. Johnson, R.~C. Rashkov and K.~S. Viswanathan,
  \emph{{Flavoured large N gauge theory in an external magnetic field}},
  \href{https://doi.org/10.1088/1126-6708/2007/10/019}{\emph{JHEP} {\bfseries
  10} (2007) 019}, [\href{https://arxiv.org/abs/hep-th/0701001}{{\ttfamily
  hep-th/0701001}}].

\bibitem{Seiberg:1999vs}
N.~Seiberg and E.~Witten, \emph{{String theory and noncommutative geometry}},
  \href{https://doi.org/10.1088/1126-6708/1999/09/032}{\emph{JHEP} {\bfseries
  09} (1999) 032}, [\href{https://arxiv.org/abs/hep-th/9908142}{{\ttfamily
  hep-th/9908142}}].

\bibitem{Nunez:2010sf}
C.~Nunez, A.~Paredes and A.~V. Ramallo, \emph{{Unquenched Flavor in the
  Gauge/Gravity Correspondence}},
  \href{https://doi.org/10.1155/2010/196714}{\emph{Adv. High Energy Phys.}
  {\bfseries 2010} (2010) 196714},
  [\href{https://arxiv.org/abs/1002.1088}{{\ttfamily 1002.1088}}].

\bibitem{Erdmenger:2005bj}
J.~Erdmenger, J.~Grosse and Z.~Guralnik, \emph{{Spectral flow on the Higgs
  branch and AdS / CFT duality}},
  \href{https://doi.org/10.1088/1126-6708/2005/06/052}{\emph{JHEP} {\bfseries
  06} (2005) 052}, [\href{https://arxiv.org/abs/hep-th/0502224}{{\ttfamily
  hep-th/0502224}}].

\end{thebibliography}\endgroup
\end{document}